\documentclass{emulateapj}
\usepackage{graphicx, rotating}

\slugcomment{Accepted 6 May 2009} 
\shorttitle{Spitzer Spectroscopy of the Antennae}
\shortauthors{Brandl et al.}
\newcommand{\2}{H{\sc ii}}
\newcommand{\Msol}{\,M$_{\odot}$}
\newcommand{\Lsol}{\,L$_{\odot}$}
\newcommand{\Zsol}{\,Z$_{\odot}$}

\begin{document}

\title{Spitzer-IRS Study of the Antennae Galaxies NGC\,4038/39}

%
%
\author{
B.R. Brandl\altaffilmark{1}, 
L. Snijders\altaffilmark{1},
M. den Brok\altaffilmark{2},
D.G. Whelan\altaffilmark{3},
B. Groves\altaffilmark{1},
P. van der Werf\altaffilmark{1},
V. Charmandaris\altaffilmark{4},
J.D. Smith\altaffilmark{5},
L. Armus\altaffilmark{6},
R.C. Kennicutt, Jr.\altaffilmark{7},
and J.R. Houck\altaffilmark{8}}

\altaffiltext{1}{Leiden Observatory, Leiden University, P.O. Box 9513, 2300
       RA Leiden, The Netherlands} 
\email{brandl@strw.leidenuniv.nl}
\altaffiltext{2}{Kapteyn Astronomical Institute, University of Groningen, 
       P.O. Box 800, 9700 AV Groningen, The Netherlands}
\altaffiltext{3}{Department of Astronomy, University of Virginia, P.O. Box 400325,
       Charlottesville, VA 22904-4325}
\altaffiltext{4}{Department of Physics, University of Crete, P.O. Box 2208
       GR-71003, Heraklion, Greece; IESL/Foundation for Research and
       Technology-Hellas, GR-71110, Heraklion, Greece; and Chercheur
       Associ\'e, Observatoire de Paris, F-75014, Paris, France}
\altaffiltext{5}{University of Toledo, 2801 W. Bancroft, Toledo, 
       OH 43606-3390, USA}
\altaffiltext{6}{Caltech, Spitzer Science Center, MS 314-6, Pasadena, 
       CA 91125, USA}
\altaffiltext{7}{Institute of Astronomy, University of Cambridge, Madingley Road, 
       Cambridge CB3 0HA, UK}
\altaffiltext{8}{Cornell University, Astronomy Department, Space Sciences Building,
       Ithaca, NY 14853, USA}

\begin{abstract}
Using the Infrared Spectrograph on the Spitzer Space Telescope, we
observed the Antennae galaxies obtaining spectral maps of the entire
central region and high signal-to-noise $5-38\mu$m spectra of the two
galactic nuclei and six infrared-luminous regions.

The total infrared luminosity of our six IR peaks plus the two nuclei
is $L_{\rm IR} = 3.8\cdot 10^{10}$\Lsol, with their derived star formation
rates ranging between 0.2 and 2\Msol/yr, with a total of 6.6\Msol/yr.
None of the typical mid-infrared tracers of AGN activity is detected
in either nucleus of the system, excluding the presence of an dust
enshrouded accretion disk.  The hardest and most luminous radiation
originates from two compact clusters in the southern part of the
overlap region, which also have the highest dust temperatures.  PAH
emission and other tracers of softer radiation are spatially extended
throughout and beyond the overlap region, but regions with harder and
intenser radiation field show a reduced PAH strength.

The strong H$_2$ emission is rather confined around the nucleus of
NGC\,4039, where shocks appear to be the dominant excitation
mechanism, and the southern part of the overlap region, where it
traces the most recent starburst activity.  The luminosity ratio
between the warm molecular gas (traced by the H$_2$ lines) and the
total far-IR emission is $\sim 1.6\cdot 10^{-4}$, similar to that
found in many starburst and ULIRGs.  The total mass of warm H$_2$ in
the Antennae is $2.5\cdot 10^7$\Msol, with a fraction of warm to total
H$_2$ gas mass of about 0.35\%.  The average warm H$_2$ temperature is
$302\pm 26$\,K and appears anti-correlated with the radiation field
hardness, possibly due to an evolution of the PDR morphology.  The
previously reported tight correlation between the H$_2$ and PAH
emission was not found but higher total PAH emission to continuum
ratios were found in PDRs with warmer gas.
\end{abstract}

\keywords{Telescopes: {\em Spitzer}; 
          Galaxies: interactions, starburst, star clusters;
	  ISM: dust, extinction, HII regions, molecular hydrogen;
	  Infrared: galaxies}


\section{Introduction}
\label{secintro}
The Antennae galaxies, NGC 4038/39 (Arp 244), are arguably the
best-known example of interacting galaxies.  The two galactic nuclei
of NGC\,4038 and NGC\,4039, and the dust-obscured overlap region in
between, exhibit one of the most stunning examples of starburst
activity\footnote{Throughout the paper we use both terms ULIRGs and
  starburst galaxies (following \citet{wee81} who coined the term
  `starburst' for the high star formation activity in NGC\,7714).
  ULIRGs, unless dominated by AGN activity, constitute the subgroup of
  the most luminous starbursts with a total infrared luminosity of
  $L_{\rm IR} > 10^{12}$\Lsol\ \citep{san96}.}  in the nearby
Universe.  They are the first system in the \citet{tom77} merger
sequence.  In the Antennae we observe the symmetric encounter between
two normal Sc type spirals at an early merger stage \citep{mih96}.

From the four {\sl IRAS} bands \citet{san03} derived total infrared
luminosities of $\log L_{\rm FIR} = L_{40-400\mu m} = 10.73$ and $\log
L_{\rm IR} = L_{8-1000\mu m} = 10.84$, corresponding to $L_{\rm FIR} =
5.6\cdot 10^{10}$\Lsol\ and $L_{\rm IR} = 7.2\cdot 10^{10}$\Lsol\ for
a distance of 22~Mpc.  With a luminosity just below
$10^{11}$\Lsol\ the Antennae does not classify as a luminous infrared
galaxy (LIRG).

The distance to the Antennae has been a subject of recent controversy.
Earlier estimates by \citet{wil00,gao01} found 20~Mpc, and
\citet{whi99} determined 19.2~Mpc.  However, using {\sl HST-ACS}
\citet{sav08} observed the tip of the red giant branch of the stellar
populations in the southern tail of the Antennae, and determined a
distance of only $13.3\pm 1.0$~Mpc.  This estimate was soon again
altered when, recently, \citet{sch08} derived $22.3\pm 2.8$~Mpc from
the type~Ia supernova 2007sr light curve, much closer to the old
estimates, which were mainly based on the Hubble flow. \citet{sch08}
also re-analyzed the {\sl HST-ACS} data and identified a different
location of the tip of the red giant branch, which agreed with their
SN1a distance.  Combining three independent methods \citet{sch08}
determined a distance of $22\pm 3.0$~Mpc, which is the distance we
assume throughout this paper.

The first deep optical images of the Antennae taken with the Wide
Field Camera on {\sl HST} revealed over 700 point-like objects
\citep{whi95}.  Subsequent V-band observations with {\sl WFPC2}
increased the sensitivity by three magnitudes and revealed between 800
and 8000 clusters in four age ranges \citep{whi99}.  These and related
observations at optical and near-IR wavelengths
\citep[e.g.,][]{fri99,whi02,kas03,bra05,men05,fal05,bas06} started
extensive studies of the properties of extragalactic star clusters in
a statistically significant way.  The Antennae provide an excellent
environment to study the transformation of supergiant molecular clouds
(SGMCs) into super star clusters (SSCs) \citep{del06} and their
further dynamical evolution \citep{whi99,fal05,bas06,and07,whi07}.
Numerous other studies \citep{gil00,men01,men02,gil07} focused on
detailed investigations of a few individual SSCs via means of near-IR
spectroscopy.

The first high signal-to-noise mid-IR images of the Antennae were made
by the Infrared Space Observatory {\sl ISO}.  {\sl ISOCAM}
observations at angular resolutions of $5-8''$ showed that the overlap
region contributes more than half of the total luminosity observed in
the $12.5-18\mu$m range \citep{vig96}.  The comparison between {\sl
  HST} and {\sl ISOCAM} images revealed a compact, optically-obscured
knot which produces 15\% of the total $12.5-18\mu$m luminosity
\citep{mir98}.  Further {\sl ISO} studies were performed by
\citet{kun96,fis96,kla97} and \citet{haa05} followed by {\sl
  Spitzer-IRAC} $3-8\mu$m observations \citep{wan04}.  Recent
sub-arcsecond imaging and N-band spectroscopy with {\sl VLT-VISIR}
resolved the \2~region / PDR interface and revealed a highly obscured
SSC that does not have an optical or near-IR counterpart
\citep{sni06}.

Observations at the longer far-infrared, sub-millimeter and radio
wavelengths generally agree with the mid-IR picture.  The early works
of \citet{hum86} with the {\sl VLA} at 1.465~GHz and 4.885~GHz found
thermal radio knots, which account for about 35\% of the total
emission, coinciding with peaks in H$_{\alpha}$ emission associated
with recent star formation.  The high resolution radio maps with the
{\sl VLA} \citep{nef00} at 6 and 4~cm became the basis for many
subsequent studies.  \citet{nik98} mapped the Antennae in the
[\ion{C}{2}] cooling line associated with photo-dissociation regions
at an angular resolution of $55''$ and found that the starburst
activity is confined to small regions of high star formation
efficiency.  \citet{wil00,wil03} found an excellent correlation
between the strengths of the CO emission and the $15\mu$m broad band
emission seen by {\sl ISO}, and determined masses of $(3-6)\cdot
10^8$\Msol\ for the largest molecular complexes, typically an order of
magnitude larger than the largest structures found in the disks of
more quiescent spiral galaxies.  The total molecular gas
mass\footnote{These estimates generally use the standard Galactic
  conversion factor of $M_{H_2} = 4.78\times \left( L_{CO}/K km s^{-1}
  pc^2 \right)$\Msol\ to derive the total H$_2$ mass from the CO
  luminosity, multiplied by 1.36 to account for the heavier elements.}
in the overlap region is $1.2\cdot 10^9$\Msol\ \citep{sta90} --
several times higher than the amount of gas in the two nuclei -- while
the molecular gas mass for the entire Antennae system is $1.5\cdot
10^{10}$\Msol\ \citep{gao01}.  At a star formation rate (SFR) of $\sim
20$\Msol\,yr$^{-1}$ \citep{zha01}, the gas consumption timescale is
only 700~Myr.  \citet{gao01} also estimated a ``normal'' star
formation efficiency (SFE) over the entire Antennae system of
$L_{\rm IR}/M_{H_2}\sim 4.2$\Lsol/\Msol, which is similar to that of GMCs
in the Galactic disk.  \citet{sch07} combined $^{12}$CO\,(1-0), (2-1),
(3-2) and $870\mu$m maps with data from X-ray to radio and PDR models,
and found that the clouds have dense ($4\times 10^4 cm^{-3}$) cores
and low kinetic temperatures ($\leq 25$\,K), showing no signs of
intense starburst activity.

Finally, numerous X-ray observations, predominantly with {\sl
  Chandra}, have revealed many X-ray sources, including several
long-term variable, ultra-luminous X-ray (ULX) sources
\citep{zez02a,zez06}.  Although most ULXs are likely black hole/high
mass X-ray binaries accreting via Roche lobe overflow, \citet{fen06}
have argued that the most luminous ULX, X-16, is a candidate for an
intermediate mass black hole (IMBH), while \citet{zez07} note that its
X-ray luminosity could be produced by a $\sim 80$\Msol\ black hole
accreting at the Eddington limit.  From the spatial distribution of
the infrared counterparts to the ULXs, \citet{cla07} concluded that
they mainly trace the recent star formation history.  \citet{bal06a}
studied metal abundances in the hot interstellar medium and found
variations from 0.2\Zsol\ to $20-30$\Zsol, but found no correlation
between the radio/optical star formation indicators and metallicity
\citep{bal06b}.

In this paper we study the central region of the Antennae galaxies
based on spatially resolved mid-IR spectroscopy.  Our aim is to
characterize the starburst activity in general, the super star
clusters in particular, and their interplay (heating and ionization)
with the interstellar medium.  From the mid-infrared spectra we derive
a total infrared luminosity for each of our six star forming peaks.
These luminosities are used to estimate the star formation rates.
Particular attention will be given to the H$_2$ rotational lines,
which probe the warm component of molecular hydrogen and which will be
used to derive the gas masses and temperatures.  The outline of this
paper is as follows: In the next section we describe the observations
along with the data reduction and calibration.  Section~\ref{secanaly}
discusses how the spectral features have been measured and the key
parameters were derived, and section~\ref{secdiscus} presents the
results, both on individual clusters and on the central region as a
whole, followed by the summary.


\section{Observations and Data Reduction}
\label{secobs}
We used the Infrared Spectrograph\footnote{The {\sl IRS} was a
collaborative venture between Cornell University and Ball Aerospace
Corporation funded by NASA through the Jet Propulsion Laboratory and
the Ames Research Center.} ({\sl IRS}) \citep{hou04} on board the
Spitzer Space Telescope \citep{wer04} to observe the central
interaction region of the Antennae galaxies at unprecedented depth.
The {\sl IRS} observations of the Antennae are part of a guaranteed
time program (PI Houck, Spitzer PID~21) on the spectroscopic study of
star formation in interacting galaxies.  The observing parameters are
listed in Table~\ref{tabobsprop} and described in more detail in
Sections~\ref{sechires} and \ref{seclores}.

\begin{deluxetable}{l r r}
\centering 
\tabletypesize{\footnotesize}
\tablecaption{Observing Parameters} 
\tablewidth{0pt} 
\tablehead{ 
\colhead{} &
\colhead{`hires'} &
\colhead{`lores'} 
}
\startdata 
Obs.date     & 3 Jan 2005 &  3 Jul 2005 \\
             &            & 12 Jul 2005 \\
             &            & 23 Jan 2006 \\
AOR\tablenotemark{a}-keys     & 3842304    & 3841536 \\
             & 3842560    & 3841792 \\
             &            & 16707840 \\
             &            & 16708096 \\
             &            & 16708352 \\
$t_{int}$\tablenotemark{b} & $24\times 30$s (SH) & $2\times 14$s (SL)\\
             & $10\times 60$s (LH) & $2\times 14$s (LL) \\
Slit sizes &  $4.\!''7\times 11.\!''3$ (SH) & $3.\!''6\times 57''$ (SL) \\
           &  $11.\!''1\times 22.\!''3$ (LH) & $10.\!''5\times 168''$ (LL) \\
\enddata
\tablenotetext{a}{Astronomical Observing Request}
\tablenotetext{b}{Number of exposures per position $\times$ exposure time}
\label{tabobsprop}
\end{deluxetable}

\subsection{`Hires' Observations}
\label{sechires}
We have made observations in {\sl IRS} high resolution (`hires') mode
at $R=600$ with the Short-High (SH) [$9.9-19.6\mu$m] and Long-High
(LH) [$18.7-37.2\mu$m] modules.  The angular slit sizes are listed
Table~\ref{tabobsprop} and correspond to approximately $500\times
1200$\,pc$^2$ (SH) and $1200\times 2380$\,pc$^2$ (LH) at the distance
of the Antennae.

\begin{figure}[tb]
\epsscale{1.1}
\plotone{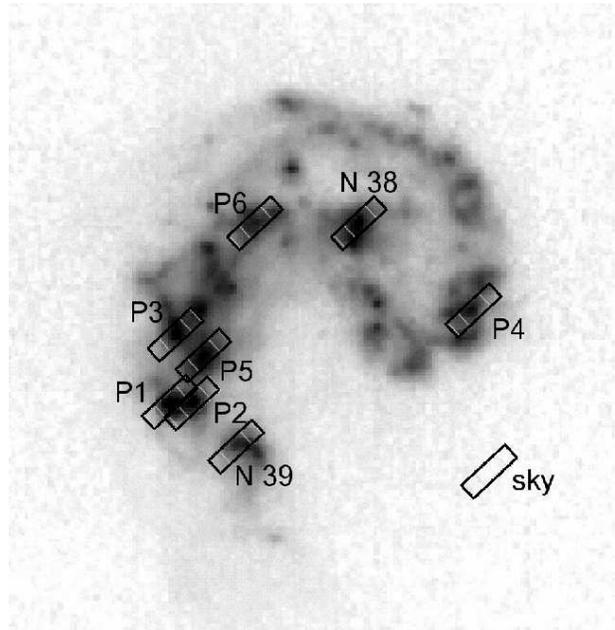}
\caption{Illustration of the {\sl IRS} SH slit positions for the given
         observing date, over-plotted on the IRAC $8\mu$m map
         \citep{wan04} of the Antennae.  North is toward the upper
         right.
         \label{figslitpos}}
\end{figure}

The spectra from the {\sl IRS} `hires' modules were taken with a
standard staring mode Astronomical Observing Template (AOT), producing
two exposures per `cycle' at separate nod positions along the slit.
The coordinates of the eight observed regions within NGC\,4038/39 are
listed in Table~\ref{tabspecpos} and the slit positions are
illustrated in Fig.~\ref{figslitpos}.  We note that the {\sl IRS}
position of the nucleus of NGC\,4038 does not exactly coincide with
the radio nucleus but is slightly shifted toward the north to also
include the nearby, active star forming region.

\begin{deluxetable*}{l r r c r r r r r}
\tabletypesize{\footnotesize}
\tablecaption{Parameters of the Pointed Observations with Cluster 
              Designations from the Literature} 
\tablewidth{0pt} 
\tablehead{ 
\colhead{Name} &
\colhead{RA (J2000)} & 
\colhead{Dec (J2000)} & 
\colhead{LH SF\tablenotemark{a}} &  
\colhead{WS95\tablenotemark{b}} &
\colhead{W99\tablenotemark{c}}&
\colhead{WZ02\tablenotemark{d}}&
\colhead{B05\tablenotemark{e}}&
\colhead{N00\tablenotemark{f}}
}
\startdata 
nucleus 4038  &  12 01 53.01 & -18 52 02.7 & 0.77 & & & & & \\
nucleus 4039  &  12 01 53.54 & -18 53 10.2 & 0.63 & & & & & \\
peak 1 &  12 01 54.98 & -18 53 05.7 & 0.65 & 80 & \nodata &  3  &  157  &  2-1 \\
peak 2 &  12 01 54.58 & -18 53 03.4 & 0.56 &  86(89/90) & 11/9/14/30/33/39 & 5 & 136 & 2-6 \\ 
peak 3 &  12 01 55.39 & -18 52 48.9 & 0.50 & 119/120(/117) & 19/17 &  7  &  176  &  4-2 \\
peak 4 &  12 01 50.42 & -18 52 12.6 & 0.59 & 405 & 2 &  25 & \nodata &  11-2 \\ 
peak 5 &  12 01 54.75 & -18 52 51.1 & 0.45 & 115 & \nodata & \nodata &  148  &  3-5 \\ 
peak 6 &  12 01 54.80 & -18 52 13.5 & 0.56 & 384/382/389 & 18 &  24 &  154  &  5-5 \\ 
reference sky &  12 01 49.00 & -18 52 50.0 & & & & & & \\
\enddata
\label{tabspecpos}
\tablenotetext{a}{Scaling factor to multiply with the LH fluxes to
                  match SH.}
\tablenotetext{b}{WS95: optical imaging with the HST WF/PC
                  \citep{whi95}.}
\tablenotetext{c}{W99: optical imaging with HST WFPC2 \citep{whi99}.}
\tablenotetext{d}{WZ02: \citet{whi02}}
\tablenotetext{e}{B05: near-IR imaging with Palomar WIRC
                  \citep{bra05}.}
\tablenotetext{f}{N00: 4 and 6~cm radio observations with the VLA
                  \citep{nef00}.}
\end{deluxetable*}

The data were pre-processed by the Spitzer Science Center data
reduction pipeline version 13.2 (Observer's Manual, 2004).  To
avoid uncertainties introduced by the flat-fielding in earlier
versions of the automated pipeline processing, we started from the
two-dimensional, unflat-fielded data products.  These products are
part of the ``basic calibrated data'' (BCD) package provided by the
Spitzer Science Center.

First, all images were cleaned of bad pixels using the IDL procedure
{\sl IRSCLEAN} with a `Maskval=128'.  The various 2-dimensional
spectra from the same nod position were median-combined within the
Spectral Modelling, Analysis, and Reduction Tool (SMART) version
5.5.1. \citep{hig04}.  A median `sky' spectrum was computed from both
nods of the sky position (\#9) and subtracted from each median nod
image.  The spectral extraction was also done within SMART, using full
aperture extraction.  The `fluxcon' tables were disabled by setting
them to `s1'.  Instead, the spectra were flat-fielded and flux
calibrated by multiplication with the relative spectral response
function (RSRF) using the {\sl IRS} RSRF of the standard star
$\xi$-Dra for both SH and LH spectra.  We have not corrected for
periodic ``detector fringing'' in the spectra since these artifacts
have no effect on the analysis carried out in this paper.

After these steps there was a noticeable and expected mismatch between
the fluxes measured by the {\sl IRS} SH and LH modules at the same
wavelengths.  This mismatch is most likely due to extended emission,
picked up by the wider LH slit.  For an ideal point source there would
be very little mismatch while for uniform surface brightness SH would
only see 21\% of the flux measured by LH.  In principle there are
three possibilities to match the two spectra:
{\sl (i)} scale LH down to match SH,
{\sl (ii)} scale SH up to match LH, or
{\sl (iii)} use a wavelength dependent scaling factor that affects
both SH and LH.  {\sl (i)} is best if the spectrum of a compact source
surrounded by extended emission is of interest, {\sl (ii)} should be
used to quantify the extended emission, and {\sl (iii)} is arguably
the best method to account for both components but requires additional
assumptions.  Since we are most interested in the spectral information
from unresolved SSCs -- which are fully covered by both SH and LH
slits -- we have chosen method {\sl (i)}.  In any case, the
quantitative results derived in this paper are from the lines covered
by the SH module and are not affected by the scaling.  The scaling
factors `SF' applied to the LH fluxes are listed in
Table~\ref{tabspecpos}.  The resulting {\sl IRS} `hires' spectra from
the two nuclei and the six infrared peaks are shown in
Fig.~\ref{figspecall}. The total observing time was 4.4~hours.

\subsection{`Lores' Observations}
\label{seclores}
\begin{figure}[tb]
\epsscale{1.1}
\plotone{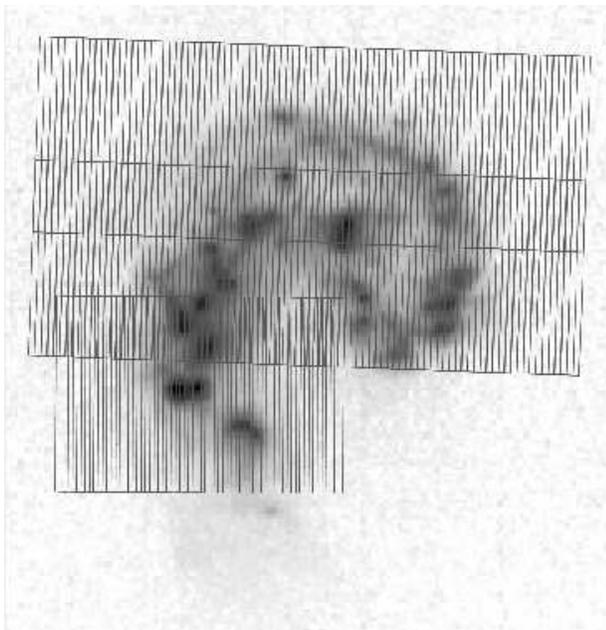}
\caption{Illustration of the areal coverage of the {\sl IRS} SL
         spectral map overlaid on the {\sl IRAC} $8\mu$m map. The
         larger part on top is the initial map while the lower,
         smaller area indicates the ``filler'' part.  North is toward
         to upper right.
         \label{figloresmap}}
\end{figure}

We have also made observations in the {\sl IRS} low resolution
(`lores') mode at $R=65-130$ with the Short-Low (SL) [$5.2-14.5\mu$m]
and Long-Low (LL) [$14.0-38.0\mu$m] modules.  The angular slit sizes
are listed Table~\ref{tabobsprop}.  The observations were made with
both `lores' modules in spectral mapping mode.  The AORs were
intentionally designed to make use of the facts that the sub-slits SL1
and SL2 are adjacent and record data simultaneously, and that the
orientation would have changed after six months (half a solar orbit).
However, an unfortunate delay of six months in the scheduling doubled
the spatial offset between SL1 and SL2 instead of providing the same
spatial coverage.  The problem was detected and corrected by
requesting another SL1 map six months later, specifically designed to
fill in the missing parts.  The mapped areas are illustrated in
Fig.~\ref{figloresmap}.  The total mapping time of all observations
combined was 7.1~hours.

All data were processed with the standard pipeline version~13. Further
reduction and analysis started from the two-dimensional, flat-fielded
data products, which are part of the ``basic calibrated data'' package
provided by the Spitzer Science Center.

Most of the reduction and spectral analysis of the large data cube has
been made with {\sl CUBISM} \citep{smi07a}.  {\sl CUBISM} is an IDL
program developed by the SINGS Legacy team to combine slit spectra at
various angles and positions and create spectral maps and local
spectra.  Artifacts in the resulting spectral maps were cleaned with a
combination of a local noise estimate and median replacement, and the
procedure {\sl IRSCLEAN}, version~1.3. The latter is based on a
multi-resolution analysis algorithm \citep{mur95}, where an image is
subsequently smoothed to different scales, using a wavelet transform.
{\sl IRSCLEAN} is a contributed {\sl IRS} software, which we modified
slightly to use a more localized noise estimate on a sub-slit basis.
We have also used Gaussian rather than uniform noise estimates.  The
algorithm was first tested on the ``sky'' spectra.
Fig.~\ref{figsixmaps} shows the resulting line maps for six important
spectral features: the fine-structure lines of [\ion{Ne}{2}] and
[\ion{S}{4}], the $11.3\mu$m and $8.6\mu$m emission feature of PAHs,
and the molecular hydrogen H$_{2}$\,S(2) and S(3) lines.


\section{Analysis}
\label{secanaly}
\begin{figure*}[hp]
\epsscale{1.1}
\plotone{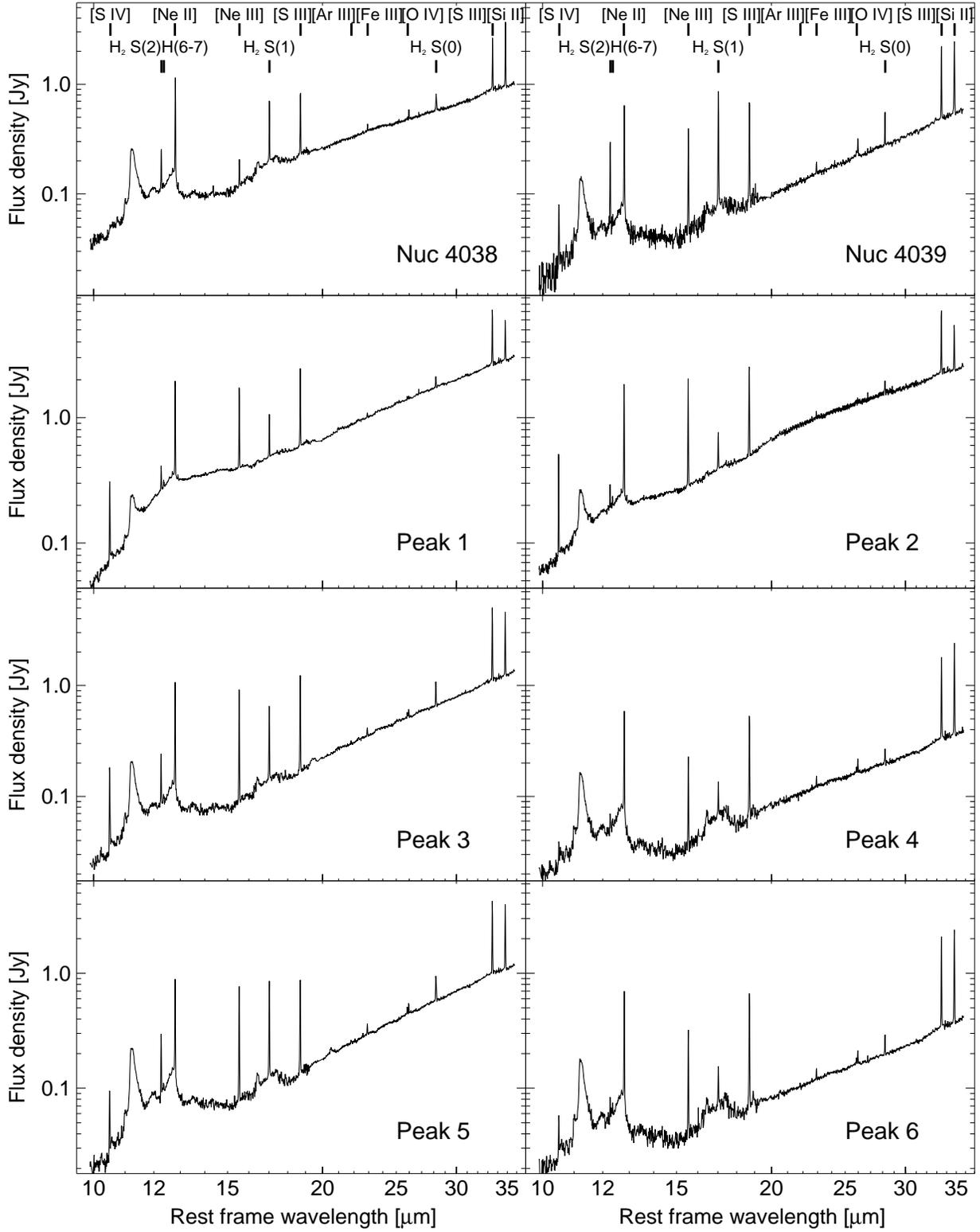}
\caption{IRS SH+LH spectra of the two nuclei of NGC\,4038 and
         NGC\,4039 and the six infrared peaks listed in
         Table~\ref{tabspecpos} and illustrated in
         Fig.~\ref{figslitpos}. Most of the detected spectral lines
         are labeled.\label{figspecall}}
\end{figure*}

Fig.~\ref{figspecall} reveals the spectral richness of the high
signal-to-noise spectra.  Important common features include the PAH
emission bands, silicate absorption features, ionic fine-structure
lines, atomic and molecular hydrogen lines, and the spectral
continuum.  In this Section we describe how the quantities relevant to
our discussion have been derived from the `lores' and `hires' spectra.
We emphasize the complementary character of the `lores' and `hires'
data in our analysis.  While the latter is used to derive accurate
line fluxes the former is mainly used for a qualitative assessment of
the spatial characteristics and for cross-checks with the `hires'
data.

\subsection{Continuum Flux Densities}
\label{seccont}
In order to characterize the slope of the spectral continuum we have
derived the flux densities $F_{\nu}$ at two narrow wavelength
intervals, namely at $14.75 - 15.25\mu$m and at $29.5 - 30.5\mu$m,
here referred to as $F_{15\mu m}$ and $F_{30\mu m}$, respectively.
These wavelengths have been chosen as they are not significantly
affected by known emission or absorption features.  Both intervals
contain about 20 resolution elements, and the continuum within that
narrow range is assumed to have a linear slope.  The flux densities of each
resolution element within the wavelength intervals have been averaged
to reduce their sensitivity to noise spikes and narrow spectral
features, resulting in the quasi-monochromatic flux densities
$F_{15\mu m}$ and $F_{30\mu m}$.

This approach is similar to the one described in \citet{bra06} and is
arguably the best direct estimate of the spectral continuum.  The flux
measurements are listed in Table~\ref{tabspeccont}.  The continuum
fluxes will be discussed in detail in Section~\ref{seccontslp}.

\begin{deluxetable}{l r r c}
\tabletypesize{\footnotesize} 
\tablecaption{Continuum Fluxes} 
\tablewidth{0pt} 
\tablehead{
\colhead{} & 
\colhead{$F_{15\mu m}$} &
\colhead{$F_{30\mu m}$} &
\colhead{$F_{15\mu m} / F_{30\mu m}$}\\
\colhead{Position} & 
\colhead{[mJy]} &
\colhead{[mJy]} & 
\colhead{}
}
\startdata 
nuc 4038 & 103 &  649 & 0.159 \\
nuc 4039 &  39 &  334 & 0.117 \\
peak 1 & 383 & 1994 & 0.192 \\
peak 2 & 260 & 1853 & 0.140 \\
peak 3 &  79 &  787 & 0.101 \\
peak 4 &  32 &  231 & 0.137 \\
peak 5 &  69 &  699 & 0.099 \\
peak 6 & 35 &  233 & 0.150 \\
\enddata
\label{tabspeccont}
\tablecomments{The flux densities listed here and in the subsequent
               tables are the values directly measured within the {\sl
               IRS} SH slit aperture of $4.\!''7\times 11.\!''3$.  The
               fluxes longward of $20\mu$m were measured with the
               larger {\sl IRS} LH slit but have been scaled down by
               the scaling factors listed in Table~\ref{tabspecpos}.}
\end{deluxetable}

\subsection{Polycyclic Aromatic Hydrocarbons (PAHs)}
\label{pahanalys}
The spectra in Fig.~\ref{figspecall} show that the spectral shape at
shorter wavelengths is largely dominated by the strong emission
features from PAHs.  The strength and equivalent width of the PAH
features were measured using the IDL program {\sl PAHFIT}
\citep{smi07b}. {\sl PAHFIT} decomposes the spectra into the
individual contributions from starlight, thermal emission from dust,
atomic and molecular emission lines, and the PAH emission bands. The
latter, combined with extinction dominated by the silicate absorption
bands around $9.7\mu$m and $18\mu$m, is the main application of the
routine and both are treated meticulously.

We have chosen to measure the below listed PAH features from the
`hires' spectra rather than the `lores' maps for two reasons.  First,
due to the longer integration times the signal-to-noise is higher than
at individual positions of the spectral map.  Second, all other
spectral diagnostics were derived from the `hires' spectra as well,
and an accurate comparison does benefit from using the same slit
apertures.  However, since {\sl PAHFIT} has been designed to analyze
{\sl IRS} `lores' spectra and does not accurately fit narrow emission
lines we removed the emission lines from the `hires' spectra for the
purpose of the PAH measurements.  This was done by linear
interpolation between the fluxes directly short-wards and long-wards
of the emission lines.  {\sl PAHFIT} uses Drude profiles to fit 15 PAH
features of different width in the {\sl IRS}-SH range.

The thermal dust continuum is fitted by a combination of modified
black bodies at fixed temperatures of 35, 40, 50, 90, 135, 200 and
300~Kelvin.  {\sl PAHFIT} can automatically correct for extinction
estimated from the depth of the $9.7\mu$m silicate absorption feature.
However, we have turned off this automatic feature since the `hires'
spectra cover only part of the $9.7\mu$m silicate absorption and we
noticed that, under these conditions, extinction is being used as a
free parameter to optimize fit results without strong physical
motivation.

The central wavelength was fixed and the full width at half maximum
(FWHM) was allowed to vary by at most 20\%.  Several adjacent, broad
PAH features cannot be resolved and are blended into one broad
complex, and we list the combined fluxes from both components.  Such
cases are the features at 11.23 and $11.33\mu$m, and at 12.62 and
$12.69\mu$m.  The PAH features at 14.04, 15.9, 18.92 and $33.1\mu$m
were not included in the analysis because of their intrinsic weakness.
The measured PAH fluxes and equivalent widths are listed in
Table~\ref{tabpahs}.  The listed uncertainties include the uncertainty
in the absolute flux calibration, which we estimate to be in the order
of 10\%, and the fit error as given by {\sl PAHFIT}.  Both
contributions have been added in quadrature.

\begin{deluxetable*}{l r r r r r r r r r}
\tabletypesize{\scriptsize} 
\tablecaption{PAH feature strengths longward of $10\mu$m from the 
              `hires' spectra}
\tablewidth{0pt} 
\tablehead{ 
\colhead{} & 
\colhead{F$_{10.7\mu m}$\tablenotemark{a}} &
\colhead{F$_{11.3\mu m}$\tablenotemark{a}} &
\colhead{F$_{12.0\mu m}$\tablenotemark{a}} &
\colhead{F$_{12.7\mu m}$\tablenotemark{a}} &
\colhead{F$_{13.5\mu m}$\tablenotemark{a}} &
\colhead{F$_{14.2\mu m}$\tablenotemark{a}} &
\colhead{F$_{16.5\mu m}$\tablenotemark{a}} &
\colhead{F$_{17.4\mu m}$\tablenotemark{a}} &
\colhead{F$_{17.9\mu m}$\tablenotemark{a}} \\
\colhead{Position} & 
\colhead{EW$_{10.7\mu m}$\tablenotemark{b}} &
\colhead{EW$_{11.3\mu m}$\tablenotemark{b}} &
\colhead{EW$_{12.0\mu m}$\tablenotemark{b}} &
\colhead{EW$_{12.7\mu m}$\tablenotemark{b}} &
\colhead{EW$_{13.5\mu m}$\tablenotemark{b}} &
\colhead{EW$_{14.2\mu m}$\tablenotemark{b}} &
\colhead{EW$_{16.5\mu m}$\tablenotemark{b}} &
\colhead{EW$_{17.4\mu m}$\tablenotemark{b}} &
\colhead{EW$_{17.9\mu m}$\tablenotemark{b}}
}
\startdata 
nuc 4038& $ 0.73 \pm 0.40  $  & $  18.02 \pm 1.40 $ & $  9.41 \pm 0.65 $  &   $   14.24 \pm 1.36  $  &  $   3.89 \pm 0.46  $  &  $   1.20 \pm 0.24 $ & $  1.27 \pm 0.20   $  &  $  6.80 \pm 1.40  $ & $ 0.46 \pm  0.63   $ \\
        & $ 0.05 \pm 0.01  $  & $  1.31 \pm 0.15  $ & $  0.67 \pm 0.08 $  &   $   0.97 \pm 0.14   $  &  $   0.24 \pm 0.04  $  &  $   0.07 \pm 0.02 $ & $  0.06 \pm 0.01   $  &  $  0.29 \pm 0.04  $ & $	0.02 \pm  0.01  $ \\
nuc 4039& $ 0.53 \pm 0.09  $  & $  9.91 \pm 0.69  $ & $  4.45 \pm 0.25 $  &   $   7.17 \pm 0.59   $  &  $   1.97 \pm 0.18  $  &  $   6.06 \pm 0.10 $ & $  0.51 \pm 0.08   $  &  $  3.53 \pm 0.36  $ & $ 0.30 \pm  0.10   $ \\
        & $ 0.09 \pm 0.02  $  & $  1.70 \pm 0.19  $ & $  0.75 \pm 0.09 $  &   $   1.18 \pm 0.16   $  &  $   0.31 \pm 0.04  $  &  $   0.09 \pm 0.02 $ & $  0.06 \pm 0.01   $  &  $  0.42 \pm 0.06  $ & $	0.03 \pm  0.01  $ \\
peak 1 & \nodata  &  $ 8.49 \pm 0.90 $  &  $ 9.83 \pm 0.98 $  &  $ 17.70 \pm 1.78 $  &  $ 15.40 \pm 1.54 $  &  $ 8.06 \pm 0.81 $ & $ 1.04 \pm 0.11 $  & $ 1.56 \pm 0.16 $  &  $ 0.92 \pm 0.09 $ \\
       & \nodata  &  $ 0.20 \pm 0.02 $  &  $ 0.21 \pm 0.02 $  &  $  0.34 \pm 0.03 $  &  $ 0.27  \pm 0.03 $  &  $ 0.13 \pm 0.01 $ & $ 0.02 \pm 0.01 $  & $ 0.02 \pm 0.01 $  &  $ 0.01 \pm 0.01 $ \\
peak 2 & $ 0.67 \pm 0.25  $  & $  14.41 \pm 1.57 $ & $  9.54 \pm 0.77 $  &   $   17.54 \pm 1.98  $  &  $   7.57 \pm 0.76  $  &  $   3.10 \pm 0.53 $ & $  0.29 \pm 0.36   $  &  \nodata & \nodata \\
        & $ 0.02 \pm 0.01  $  & $  0.47 \pm 0.02  $ & $  0.28 \pm 0.04 $  &   $   0.48 \pm 0.08   $  &  $   0.19 \pm 0.03  $  &  $   0.07 \pm 0.01 $ & $  0.01 \pm 0.01   $  &  \nodata & \nodata \\
peak 3 & $ 0.40 \pm 0.11  $  & $  13.67 \pm 0.69 $ & $  6.42 \pm 0.34 $  &   $   11.09 \pm 0.95  $  &  $   3.30 \pm 0.28  $  &  $   0.90 \pm 0.19 $ & $  0.90 \pm 0.15   $  &  $  3.39 \pm 0.95  $ & \nodata \\
        & $ 0.04 \pm 0.01  $  & $  1.24 \pm 0.14  $ & $  0.56 \pm 0.06 $  &   $   0.92 \pm 0.12   $  &  $   0.26 \pm 0.03  $  &  $   0.07 \pm 0.02 $ & $  0.05 \pm 0.01   $  &  $  0.18 \pm 0.04  $ & \nodata \\
peak 4 & $ 0.68 \pm 0.09  $  & $  12.17 \pm 0.82 $ & $  4.02 \pm 0.56 $  &   $   8.12 \pm 1.07   $  &  $   1.42 \pm 0.29  $  &  $   0.67 \pm 0.13 $ & $  0.58 \pm 0.07   $  &  $  3.43 \pm 0.40  $ & \nodata \\ 
        & $ 0.10 \pm 0.02  $  & $  2.06 \pm 0.23  $ & $  0.75 \pm 0.09 $  &   $   1.60 \pm 0.24   $  &  $   0.28 \pm 0.04  $  &  $   0.13 \pm 0.03 $ & $  0.09 \pm 0.01   $  &  $  0.49 \pm 0.07  $ & \nodata \\
peak 5 & $ 0.36 \pm 0.19  $  & $  15.57 \pm 0.80 $ & $  7.06 \pm 0.84 $  &   $   14.25 \pm 1.11  $  &  $   4.07 \pm 0.28  $  &  $   1.46 \pm 0.18 $ & $  0.90 \pm 0.13   $  &  $  4.03 \pm 0.78  $ & \nodata \\
        & $ 0.04 \pm 0.01  $  & $  1.79 \pm 0.21  $ & $  0.75 \pm 0.08 $  &   $   1.40 \pm 0.18   $  &  $   0.37 \pm 0.05  $  &  $   0.13 \pm 0.02 $ & $  0.06 \pm 0.01   $  &  $  0.27 \pm 0.05  $ & \nodata \\
peak 6  & $ 0.70 \pm 0.08  $  & $  12.97 \pm 0.85 $ & $  4.18 \pm 0.58 $  &   $   9.00  \pm1.17   $  &  $   1.66 \pm 0.30  $  &  $   0.96 \pm 0.11 $ & $  0.79 \pm 0.09   $  &  $  3.67 \pm 0.77  $ & \nodata \\
        & $ 0.10 \pm 0.02  $  & $  2.00 \pm 0.23  $ & $  0.70 \pm 0.08 $  &   $   1.56 \pm 0.23   $  &  $   0.29 \pm 0.04  $  &  $   0.16 \pm 0.02 $ & $  0.11 \pm 0.02   $  &  $  0.50 \pm 0.05  $ & \nodata \\
\enddata
\label{tabpahs}
\tablecomments{The listed measurements have not been corrected for 
               extinction. See comment to Table~\ref{tabspeccont} for 
               slit aperture sizes.}
\tablenotetext{a}{Integrated flux in units of $10^{-20}$W\,cm$^{-2}$}
\tablenotetext{b}{Equivalent width in units of $\mu$m}
\end{deluxetable*}

The PAH strength in the `lores' spectra has been visualized and
analyzed within {\sl CUBISM} in a more qualitative way.  We co-added
the emission from the spectral elements covered under the emission
feature and subtracted the continuum flux.  The latter was derived
from a linear interpolation between the continuum fluxes to both sides
of the feature, which were each computed from the median of typically
three spectral resolution elements.

\begin{figure*}[htb]
\epsscale{1.1}
\plotone{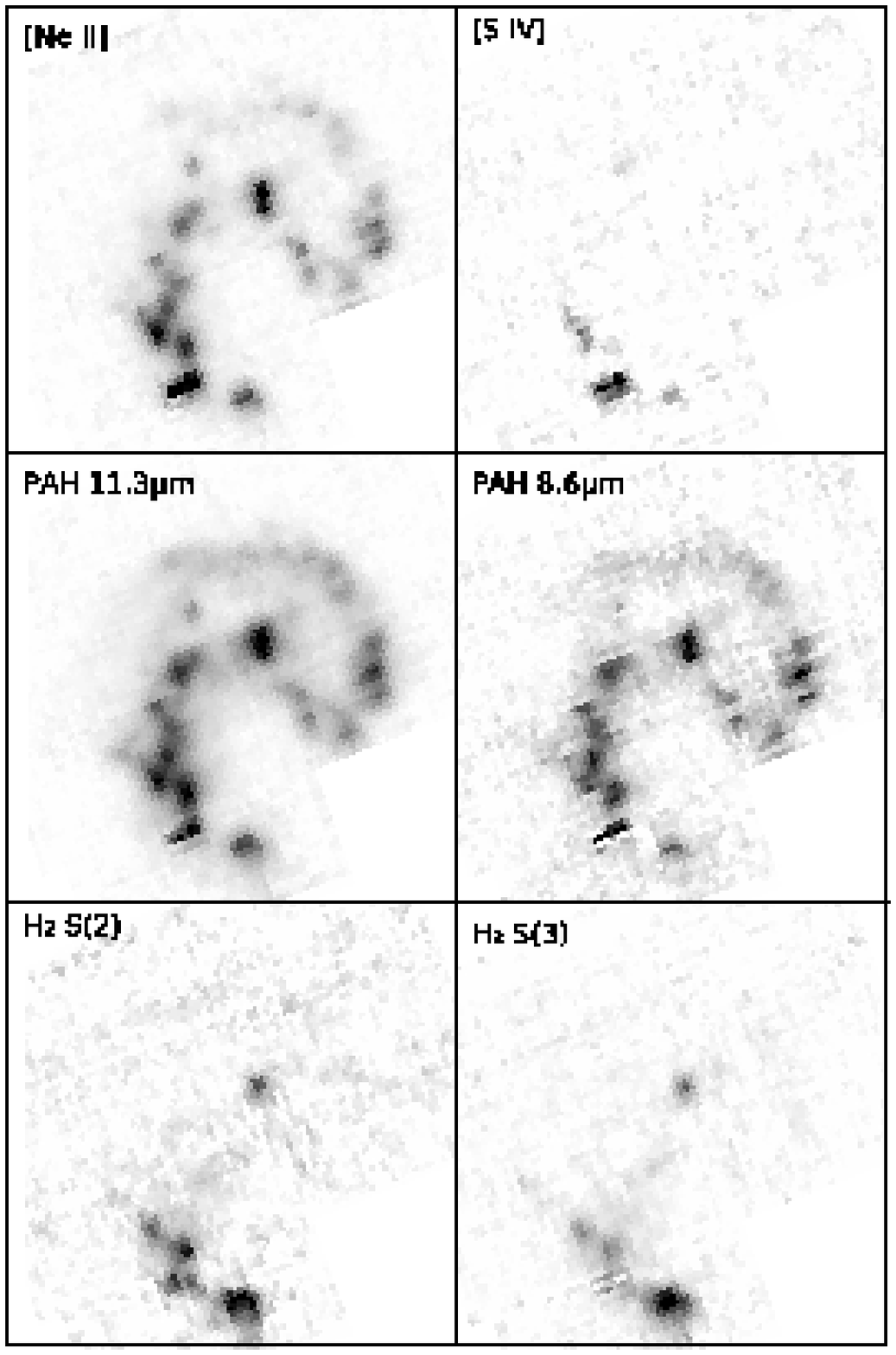}
\caption{{\sl IRS}-SL spectral maps at the [\ion{Ne}{2}] and
         [\ion{S}{4}] lines (top), the $11.3\mu$m and $8.6\mu$m PAH
         features (center), and the S(2) and S(3) rotational lines of
         H$_{2}$.  The spectral continuum has been subtracted and the
         images shown represent the pure line intensities.  The maps
         are shown in linear scaling.  North is up, east is left.  The
         images are approximately $135''$ (14.4~kpc) aside.  For a
         discussion see Section~\ref{secpropism}\label{figsixmaps}}
\end{figure*}

We produced spectral maps in the $6.2\mu$m, $8.6\mu$m and $11.3\mu$m
PAH emission features.  The former closely resembles the one at
$8.6\mu$m and is not shown in this paper.  Since the $12.7\mu$m PAH
map is affected by the [\ion{Ne}{2}] line (see Section~\ref{seclines})
we did not include it in the further analysis.  The $11.3\mu$m feature
is the one most affected by silicate absorption.  At low spectral
resolution the $8.6\mu$m map has significant contamination from the
broad $7.7\mu$m PAH feature \citep{smi07b}.  However, since both
species are similar in nature this will not affect the qualitative
conclusions drawn from the spectral maps.  The $8.6\mu$m and
$11.3\mu$m PAH maps are shown in Fig.~\ref{figsixmaps}.

\subsection{Fine-structure and Hydrogen Emission Lines }
\label{seclines}
As can be seen in Fig.~\ref{figspecall} the mid-IR wavelength range
contains numerous strong fine-structure emission lines, most
prominently from sulphur and neon.  We measured the line fluxes from
the `hires' spectra using the Gaussian line fitting tool in {\sl
SMART} and a linear baseline fit.  The results are listed in
Table~\ref{tabfslines}.  The quoted uncertainties include the
uncertainty in the absolute flux calibration, which we estimate to be
of the order of 10\%, and the fit errors provided by {\sl SMART}.
Both contributions have been added in quadrature.

\begin{deluxetable*}{l r r r r r r r r r r r}
\tabletypesize{\scriptsize}
\tablecaption{Fine-structure line fluxes in units of
$10^{-20}$W\,cm$^{-2}$}
\tablewidth{0pt}
\tablehead{
\colhead{Line} &
\colhead{[\ion{S}{4}]} &
\colhead{[\ion{Ne}{2}]} &
\colhead{[\ion{Ne}{3}]} &
\colhead{[\ion{S}{3}]} &
\colhead{[\ion{Ar}{3}]} &
\colhead{[\ion{Fe}{3}]} &
\colhead{[\ion{O}{4}]} &
\colhead{[\ion{Fe}{2}]} &
\colhead{[\ion{S}{3}]} &
\colhead{[\ion{Si}{2}]} \\
\colhead{Wavelength} &
\colhead{$10.51\mu$m} &
\colhead{$12.81\mu$m} &
\colhead{$15.56\mu$m} &
\colhead{$18.71\mu$m} &
\colhead{$21.83\mu$m} &
\colhead{$22.93\mu$m} &
\colhead{$25.89\mu$m} &
\colhead{$25.99\mu$m} &
\colhead{$33.48\mu$m} &
\colhead{$34.82\mu$m} \\
\colhead{EP\tablenotemark{a}} &
\colhead{34.97 eV} &
\colhead{21.56 eV} &
\colhead{40.96 eV} &
\colhead{23.34 eV} &
\colhead{27.63 eV} &
\colhead{16.19 eV} &
\colhead{54.93 eV} &
\colhead{ 7.90 eV} &
\colhead{23.34 eV} &
\colhead{ 8.15 eV} 
}
\startdata
nuc 4038 & \nodata & $ 3.83 \pm 0.40 $           & $ 0.29 \pm 0.03 $ & $ 
1.89 \pm 0.20 $ & \nodata           & $ 0.12 \pm 0.02 $ & $ 0.12 \pm 0.04
$ & $ 0.21 \pm 0.03 $ & $ 2.75 \pm 0.28 $ & $ 3.88 \pm 0.39 $  \\
nuc 4039 & $ 0.34 \pm 0.04 $ & $ 2.41 \pm 0.25 $ & $ 1.14 \pm 0.12 $ & $
1.91 \pm 0.20 $ & \nodata           & $ 0.11 \pm 0.01 $ & $ 0.10 \pm 0.69
$ & $ 0.23 \pm 0.31 $ & $ 2.67 \pm 0.27 $ & $ 3.25 \pm 0.33 $  \\
peak 1 & $ 1.25 \pm 0.14 $ & $ 7.05 \pm 0.77 $   & $ 4.44 \pm 0.46 $ & $
5.14 \pm 0.52 $ & $ 0.05 \pm 0.08 $ & $ 0.22 \pm 0.04 $ & $ 0.14 \pm 0.03
$ & $ 0.16 \pm 0.04 $ & $ 7.64 \pm 0.80 $ & $ 4.73 \pm 0.51 $  \\
peak 2 & $ 1.87 \pm 0.19 $ & $ 6.64 \pm 0.70 $   & $ 4.86 \pm 0.49 $ & $
6.22 \pm 0.69 $ & \nodata           & $ 0.30 \pm 0.09 $ & \nodata
& $ 0.26 \pm 0.14 $ & $ 8.72 \pm 0.93 $ & $ 5.19 \pm 0.54 $  \\
peak 3 & $ 0.78 \pm 0.08 $ & $ 3.98 \pm 0.41 $   & $ 2.48 \pm 0.26 $ & $
3.21 \pm 0.32 $ & \nodata           & $ 0.18 \pm 0.02 $ & $ 0.13 \pm 0.02
$ & $ 0.21 \pm 0.03 $ & $ 6.09 \pm 0.62 $ & $ 4.90 \pm 0.50 $  \\
peak 4 & $ 0.05 \pm 0.01 $ & $ 2.11 \pm 0.25 $   & $ 0.65 \pm 0.07 $ & $
1.37 \pm 0.14 $ & \nodata           & $ 0.06 \pm 0.01 $ & $ 0.05 \pm 0.01
$ & $ 0.10 \pm 0.01 $ & $ 3.18 \pm 0.32 $ & $ 3.18 \pm 0.32 $  \\
peak 5 & $ 0.39 \pm 0.04 $ & $ 3.33 \pm 0.34 $ & $ 2.31 \pm 0.23 $ & $
2.43 \pm 0.25 $ & $ 0.05 \pm 0.02 $ & $ 0.17 \pm 0.02 $ & $ 0.17 \pm 0.04
$ & $ 0.22 \pm 0.02 $ & $ 5.14 \pm 0.53 $ & $ 4.51 \pm 0.46 $  \\
peak 6 & $ 0.14 \pm 0.02 $ & $ 2.47 \pm 0.27 $     & $ 0.95 \pm 0.10 $ & $
1.71 \pm 0.18 $ & $ 0.02 \pm 0.00 $ & $ 0.07 \pm 0.01 $ & $ 0.06 \pm 0.01
$ & $ 0.10 \pm 0.01 $ & $ 2.55 \pm 0.26 $ & $ 3.23 \pm 0.33 $  \\
\enddata
\label{tabfslines}
\tablecomments{The listed measurements have not been corrected for 
               extinction. See comment to Table~\ref{tabspeccont} for 
               slit aperture sizes.}
\tablenotetext{a}{Excitation potential to create this ion.}
\end{deluxetable*}

Our measurements of [\ion{Ne}{3}]\,/\,[\ion{Ne}{2}] are in the range
of $0.08-0.75$, and hence lower than the
[\ion{Ne}{3}]\,/\,[\ion{Ne}{2}] $\sim 0.8$ derived by \citet{kun96}
with {\sl ISO-SWS} spectroscopy for the overlap region.  However, our
``line-luminosity weighted'' average for the peaks~1, 2, 3, 5, and 6
together is $0.57^{+0.13}_{-0.11}$, still lower than but close to the
{\sl ISO-SWS} value of 0.8.

Fig.~\ref{figspecall} also shows the lowest pure rotational
transitions of molecular hydrogen, namely the 0-0~S(0) $28.21\mu$m,
the 0-0~S(1) $17.03\mu$m, and the 0-0~S(2) $12.28\mu$m lines.  In
addition we find the atomic hydrogen lines of Humphreys-$\alpha$
H\,(7-6) $12.37\mu$m and the H\,I~(8-7) $19.06\mu$m.  The latter,
however, has only been detected in the nucleus of NGC\,4039.  The line
fluxes are listed in Table~\ref{tabh2lines}.  The quoted uncertainties
include the uncertainty in the absolute flux calibration, which we
estimate to be in the order of 10\% for the `hires' spectra and 20\%
for the spectral maps, and the fit errors provided by {\sl SMART},
again added in quadrature.

\begin{deluxetable*}{l r r r r r r r}
\tabletypesize{\footnotesize}
\tablecaption{Line fluxes of molecular and atomic hydrogen in units of
               $10^{-20}$W\,cm$^{-2}$ observed in the {\sl IRS} SL and
               SH modules.}
\tablewidth{0pt}
\tablehead{
\colhead{Transition} &
\colhead{H$_{2}$S(3)\tablenotemark{a}} &
\colhead{H$_{2}$S(2)\tablenotemark{a}} &
\colhead{H$_{2}$S(2)\tablenotemark{b}} &
\colhead{H$_{2}$S(1)\tablenotemark{b}} &
\colhead{H$_{2}$S(0)\tablenotemark{b}} &
\colhead{H (7-6)\tablenotemark{b}} &
\colhead{H (8-7)\tablenotemark{b}} \\
\colhead{Wavelength} &
\colhead{$9.67\mu$m} &
\colhead{$12.28\mu$m} &
\colhead{$12.28\mu$m} &
\colhead{$17.03\mu$m} &
\colhead{$28.22\mu$m} &
\colhead{$12.37\mu$m} &
\colhead{$19.06\mu$m} \\
\colhead{$E_l$\tablenotemark{c}} &
\colhead{1015 K} &
\colhead{ 510 K} &
\colhead{ 510 K} &
\colhead{ 170 K} &
\colhead{   0 K} &
\colhead{} &
\colhead{} 
}
\startdata
nuc 4038 & $ 0.71 \pm 0.15 $ & $ 0.52 \pm 0.11 $ & $ 0.64 \pm 0.08 $ & $ 1.43 \pm 0.15 $ & $ 0.72 \pm 0.21 $ & $ 0.06 \pm 0.01 $ & \nodata  \\
nuc 4039 & $ 3.76 \pm 0.76 $ & $ 1.99 \pm 0.42 $ & $ 1.11 \pm 0.12 $ & $ 2.37 \pm 0.24 $ & $ 0.43 \pm 0.08 $ & $ 0.04 \pm 0.05 $ & $ 0.05 \pm 0.01 $  \\
peak 1 & $\le 0.50\tablenotemark{d} $ & $0.39 \pm 0.09 $  & $ 0.62 \pm 0.08 $ & $ 1.64 \pm 0.17 $ & $ 0.81 \pm 0.09 $ & $ 0.15 \pm 0.02 $ & \nodata  \\
peak 2 & $ 0.43 \pm 0.11 $ & $ 0.45 \pm 0.11 $ & $ 0.36 \pm 0.06 $ & $ 1.23 \pm 0.13 $ & $ 1.09 \pm 0.24 $ & $ 0.11 \pm 0.04 $ & \nodata  \\
peak 3 & $ 0.53 \pm 0.12 $ & $ 0.45 \pm 0.10 $ & $ 0.66 \pm 0.07 $ & $ 1.77 \pm 0.18 $ & $ 0.90 \pm 0.10 $ & $ 0.11 \pm 0.03 $ & \nodata  \\
peak 4 & \nodata           & \nodata           & $ 0.11 \pm 0.03 $ & $ 0.19 \pm 0.02 $ & $ 0.22 \pm 0.05 $ & $ 0.06 \pm 0.02 $ & \nodata  \\
peak 5 & $ 0.69 \pm 0.14 $ & $ 0.67 \pm 0.15 $ & $ 0.95 \pm 0.11 $ & $ 2.38 \pm 0.24 $ & $ 0.87 \pm 0.10 $ & $ 0.11 \pm 0.02 $ & \nodata  \\
peak 6 & \nodata           & \nodata           & $ 0.13 \pm 0.03 $ & $ 0.24 \pm 0.03 $ & $ 0.16 \pm 0.02 $ & $ 0.05 \pm 0.01 $ & \nodata  \\
\enddata
\label{tabh2lines}
\tablecomments{The listed measurements have not been corrected for
               extinction. See comment to Table~\ref{tabspeccont} for
               the `hires' slit sizes and Section~\ref{seclines} for
               the `lores' apertures.}
\tablenotetext{a}{Observed with the {\sl IRS}-SL module. The
                  quoted uncertainties only include the fitting and 
                  photometric uncertainties but no systematic errors 
                  of the spectral mapping.}
\tablenotetext{b}{Observed with the {\sl IRS}-SH module.}
\tablenotetext{c}{Energy of lower level in temperature equivalents.}
\tablenotetext{d}{Upper limit only, due to an artifact in the spectral map.}
\end{deluxetable*}

The line strength in the `lores' spectra has been visualized and
analyzed with {\sl CUBISM} in the same way as the PAH features before.
We concentrate the analysis on the strongest fine-structure and
hydrogen lines within the range of the {\sl IRS}-SL module to have
similar spatial resolution for all maps (the slit width increases by
a factor of two long-ward of $15\mu$m).  Hence, we present the [\ion{S}{4}]
map in Fig.~\ref{figsixmaps} instead of [\ion{Ne}{3}] which lies outside
the range covered by the {\sl IRS}-SL module.  

We note that in `lores' mode, the strong, adjacent emission features of
[\ion{Ne}{2}] at $12.81\mu$m and PAH at $12.7\mu$m cannot be fully
spectrally resolved.  From the `hires' spectra we know that, on
average, the luminosity in the PAH feature is about three times higher
and thus a serious potential contaminant.  A cross-contamination may
lead to an over-estimation of the line flux, but also to an
over-estimation of the continuum flux.  For that reason we have
excluded the $12.7\mu$m PAH feature from our analysis.  A comparison
of the [\ion{Ne}{2}] line flux in the nucleus of NGC\,4039 with the
`hires' spectrum, where the components are clearly resolved, indicates
that the PAH contamination to the line flux is small ($\sim 20$\%),
and that the [\ion{Ne}{2}] map is suitable for qualitative arguments. 

The 0-0~S(1)$17.03\mu$m, the 0-0~S(2)$12.28\mu$m and the
0-0~S(3)$9.66\mu$m emission lines of molecular hydrogen were also
detected in the `lores' spectral maps.  The S(2) and S(3) lines were
both measured with the IRS-SL1 module, while the S(1) line was
measured with the IRS-LL2 module at $\sim 2\times$ lower spatial
resolution.  Unlike the fixed `hires' slit sizes and orientation, the
apertures for the spectral maps, from which the `lores' fluxes were
extracted, were best matched to the size of the emitting region and
vary from $5\times 4$ pixels ($960\times 768$\,pc) to $10\times 11$
pixels ($1.92\times 2.11$\,kpc) for the area around the nucleus of
NGC\,4039.  The fluxes of the S(2) and S(3) lines have been added to
Table~\ref{tabh2lines}.  Fig.~\ref{figsixmaps} shows the spectral maps
at the [\ion{Ne}{2}], [\ion{S}{4}], and the
H$_{2}$\,0-0~S(2)$12.28\mu$m and S(3)$9.66\mu$m lines.


\subsection{Deriving H$_2$ Temperatures and Masses}
\label{secH2ana}
The H$_{2}$ lines serve as crucial diagnostics of the conditions of
the ISM.  The H$_2$ temperatures were calculated independently for the
`lores' and the `hires' data as described in more detail in the
Appendix and are listed in Table~\ref{tablumis}.  Values range from
approximately 270\,K to 370\,K.

However Table~\ref{tablumis} does not list the uncertainties in the
temperature and mass estimates because they are dominated by three
components, which we discuss here.  First are the errors in the line
flux measurements, listed in Table~\ref{tabfslines}.  These are
typically about 3\%.  Second, the different clusters suffer different
amounts of extinction.  Correcting for extinction has two effects: it
increases the line fluxes and it changes the line ratios.  For
instance, a ``typical'' optical depth of $\tau_{9.7} = 0.2$ would
raise the temperature determined from the S(2) and S(3) lines by about
4\% ($\sim 12$\,K) since the S(3) is located in the center of the
silicate band and suffers most from extinction.  If we use the S(1)
and S(2) lines instead, the temperatures rises only by 0.7\% ($\sim
2$\,K).  The effect of extinction on the mass estimate is more
complicated since it increases linearly with the corrected line flux
but depends also on the temperature.  For $\tau_{9.7} = 0.2$ the H$_2$
mass increases by approximately 5\%.  Given the large uncertainties in
the $A_V$ or $\tau_{9.7}$ for the clusters, as discussed in
Section~\ref{secextinct} and summarized in Table~\ref{tabext}, we
decided not to correct our tabulated estimates for extinction.  Third,
aperture effects play an important role.  The `hires' aperture is
defined by the slit width and length and not matched to the size of
the H$_2$ emitting region.  For the `lores' values the aperture was
matched to the size of the region (Section~\ref{seclines}).  This
systematic error is reflected in Table~\ref{tablumis} by the
difference between the corresponding mass estimates, and is the
dominating uncertainty.

We have also calculated the H$_2$ masses following the procedure
outlined in the Appendix.  The derived masses are listed independently
for the `lores' and the `hires' data in Table~\ref{tablumis}.  Again,
the two independent mass estimates using different lines and
measurements from different instruments agree within $\sim 15$~\%
(except for peak~3), which provides confidence in the reliability of
our measurements.  A discussion of the derived H$_2$ masses is given
in Section~\ref{secH2mass}.


\subsection{Extinction Estimates}
\label{secextinct}
The wavelength range covered by the {\sl IRS-}SL and SH modules
includes two important diagnostics to estimate the amount of dust: a
broad Si=O stretching resonance, peaking at $9.7\mu$m, and an even
broader O-Si-O bending mode resonance, peaking at $18.5\mu$m.  We have
estimated the extinction using the apparent optical depth in the
$9.7\mu$m silicate feature.  We assume extinction from a foreground
screen, which is certainly an over-simplification.  However, unlike
for starburst galaxies as a whole, it is a reasonable approximation
for the extinction toward individual, embedded SSCs, where the
luminosity is provided by a central stellar cluster which is
surrounded by dust and gas clouds and PDRs.

The depth of the $9.7\mu$m silicate feature was directly measured as
the natural logarithm of the ratio of observed flux to the nominal
mid-IR continuum at $9.7\mu$m.  The latter can be best estimated (for
PAH dominated spectra) by a power law fit anchored at $5.5\mu$m and
$14.5\mu$m.  The method is described and illustrated in Fig.~2 of
\citet{spo07}.  Our estimates of the optical depth are listed in the
right column of Table~\ref{tabext}.

\begin{deluxetable}{l c c c c}
\centering 
\tabletypesize{\scriptsize} 
\tablecaption{Extinction estimates}
\tablewidth{0pt} 
\tablehead{
\colhead{Position} &
\colhead{A$_V$\tablenotemark{a}} &
\colhead{A$_V$\tablenotemark{b}} &
\colhead{A$_V$} &
\colhead{$\tau_{9.7}$\tablenotemark{e}} 
}
\startdata 
peak 1  &    4.23 & $6.2 \pm 0.3$  & $4.3\pm 0.3$\tablenotemark{c}  & 0.19 \\
peak 2  &    0.18 & $0.7 \pm 0.1$  & $1.4\pm 0.3$\tablenotemark{c}  & 0.13 \\ 
peak 3  &    0.14 & $4.8 \pm 0.4$  & \nodata                        & \nodata \\
peak 4  & \nodata & \nodata        & $0.6\pm 0.3$\tablenotemark{c}  & \nodata \\ 
peak 5  &   11.81 & $10.3 \pm 0.5$ & \nodata                        & 1.03 \\ 
peak 6  & \nodata & \nodata        & 0.72\tablenotemark{d}          & 0.05 \\ 
\enddata
\label{tabext}
\tablenotetext{a}{A$_V$ from a comparison of near-IR photometry of
                  various broad- and narrow-band fluxes with
                  {\sl STARBURST99} \citep{men05}.}
\tablenotetext{b}{A$_V$ from Br$_{\gamma}$/Pa$_{\beta}$ and the Brackett
                  series \citep{sni07b}.}
\tablenotetext{c}{A$_V$ from Br$_{\gamma}$/H$_{\alpha}$ \citep{men01}.}
\tablenotetext{d}{A$_V$ from H$_{\alpha}$/H$_{\beta}$ \citep{bas06}.}
\tablenotetext{e}{$\tau_{9.7}$ estimates from baseline fitting
                  (Section~\ref{secextinct}).}
\end{deluxetable}

We disabled the {\sl PAHFIT} default setting to automatically correct
the measurements for extinction, and hence the values in
Table~\ref{tabpahs} are the uncorrected, directly measured fluxes.
This has been done for three reasons.  First, {\sl PAHFIT} has many
free parameters, and given the limited spectral range of the {\sl
IRS}-SH module, which does not fully cover the $9.7\mu$m silicate
band, extinction could be used as yet another free parameter to
optimize the overall fit.  Second, we do not, a priori, know whether a
dust screen or mixed geometry provides a more realistic description.
Third, the extinction at mid-IR wavelengths is generally rather low in
these objects.

For a large sample of 172 near-IR sources in the Antennae,
\citet{bra05} found that the average extinction is about $A_V\sim 2$,
while the reddest clusters may be attenuated by up to 10 magnitudes.
From {\sl ISO-SWS} spectroscopy of the hydrogen recombination lines
Br$_{\gamma}$ and Br$_{\alpha}$ in the overlap region \citet{kun96}
derived an extinction of $A_V\sim 70$ (mixed case) or $A_V\sim 15$
(screen extinction).  The literature values are compared with our
measurements in Table~\ref{tabext}.

The conversion from $\tau_{9.7}$ to $A_V$ strongly depends on
the assumed extinction law.  Toward the Galactic Center there are
various estimates: $A_V/\tau_{9.7}\sim 6.7$ \citep{mon01},
$A_V/\tau_{9.7}\sim 7.8$ \citep{lut99}, and $A_V/\tau_{9.7}\sim 9.0$
\citep{mat90}.  For the Solar neighbourhood \citet{mat90} derived
$A_V/\tau_{9.7}\sim 18.5$, and \citet{dra89} calculated
$A_V/\tau_{9.7}\sim 18.1$.  Given this broad range of conversion
factors, we find that the estimates for $\tau_{9.7}$ from baseline
fitting agree reasonably well with the other values listed in
Table~\ref{tabext}.

We note that the $\tau_{9.7}$ values indicate less extinction than
derived from observations at shorter wavelengths, and the ratio of
$A_V / \tau_{9.7}$ varies significantly from source to source.  This
may be caused by a dilution effect due to the large area covered by
the {\sl IRS} slits.  An extended component of strong mid-infrared
continuum emission will decrease the relative depth of the silicate
absorption feature.  This effect is likely most predominant for peak~1.
Furthermore, optical methods mainly probe the absorption by
graphite-based dust particles, while the mid-IR methods are mainly
sensitive to distinct silicate absorption.  In the diffuse ISM
graphite and silicate-based dust are uniformly mixed and tightly
correlated \citep{roc84}.  Recently, \citet{chi07} found that in dense
clouds with $A_V\ge 12$ the linear relation between optical and mid-IR
estimates breaks down and $\tau_{9.7}$ underestimates the real amount
of dust.  However, most of our regions are below $A_V\sim 12$ and the
discrepancy between the extinction values in Table~\ref{tabext} is
primarily given by systematic uncertainties rather than ISM chemistry.


\section{Results and Discussion} 
\label{secdiscus}
In this Section we discuss and interpret our observational findings
from both the spectral maps and the spectra of the two nuclei and six
infrared peaks.  Our aim is to characterize the conditions under which
SSCs form and evolve, their properties, and how their presence affects
the surrounding interstellar medium.  We start with a discussion of
the individual regions, discuss qualitatively the ISM properties
across the central region of the Antennae, investigate dust
temperatures and radiation field, and derive cluster masses and star
formation rates.  We discuss the strength and variability of the PAH
features and focus specifically on the strength of the H$_{2}$
emission and the temperature and excitation of the molecular hydrogen
in the Antennae.


\subsection{The Nuclei of NGC\,4038 and NGC\,4039}
\label{secnuclei}
Based on sub-arcsecond near-IR spectroscopy from Keck, \citet{gil00}
found that the spectrum of the nucleus of NGC\,4039 is marked by a
strong stellar continuum and bright, extended H$_{2}$ emission.  These
authors also found strong photospheric absorption of \ion{Mg}{1},
\ion{Na}{1}, and \ion{Ca}{1} as well as from the CO $\Delta \nu = 2$
band head absorption, indicating that the continuum is dominated by
old giants and red super-giants. The Br$_{\gamma}$ line, an indicator of
recent massive star formation, was not detected.  \citet{men01} found
starburst activity a couple of arcseconds north of the nucleus of
NGC\,4038, consistent with an age of around 6 Myr.  For the two nuclei
\citet{men01} estimated an age of $65\pm15$~Myr from CO absorption.
It is clear that two galactic nuclei have lower star formation
activity than the overlap region \citep{wan04}.

Chandra detected several X-ray sources near the radio positions of the
nuclei of NGC\,4038 and NGC\,4039 \citep{zez02a}. The X-ray sources
identified with the nuclei are both luminous and spatially extended.
The NGC\,4038 nucleus spectrum is very soft and steep, and consistent
with thermal emission by a supernova-driven super-wind, possibly with
some contribution by X-ray binaries \citep{zez02a,zez02b}.  The
nucleus of NGC\,4039 is more luminous in X-rays, but its spectrum
also suggests a combination of X-ray binaries and compact supernova
remnants \citep{zez02a,zez02b}. The NGC\,4039 nucleus also has a very
steep radio spectrum \citep{nef00} suggesting that its radio emission
is non-thermal, arising from supernova remnants.  These observations
are consistent with the intermediate starburst age of about 65~Myr
estimated by \citet{men01}.  The supernovae may be responsible for
heating the ISM in the nuclear region to the observed X-ray
temperatures, and the X-ray emission, and shocks arising from the
supernovae may also be responsible for the high-excitation line
emission.  At any rate, neither X-ray nor radio observations have
provided strong evidence for activity due to black hole accretion in
either nucleus.

The {\sl IRS} spectra of the two nuclei are quite similar in terms
of their PAH strength and silicate absorption (Fig.~\ref{figspecall}).
However they differ significantly regarding their mid-IR fluxes,
20-30$\mu$m slopes, and fine-structure lines.  As mentioned in
Section~\ref{sechires} the observed position of the nucleus of
NGC\,4038 is slightly offset to the north of the radio nucleus, and
the wide {\sl IRS}-SH slit includes the nearby star forming region as
well.  Hence, the detected mid-IR emission is likely affected by the
starburst activity to the north of the nucleus of NGC\,4038.  The
nucleus and the nearby starburst region can be seen in most of the
spectral maps in Fig.~\ref{figsixmaps} as an extended structure with
two separate components.  
The strong high excitation lines in the NGC\,4039 nucleus pose
somewhat more of a puzzle.  The [\ion{Ne}{3}]/\,[\ion{Ne}{2}] ratio is
almost six times higher in NGC\,4039, and the [\ion{S}{4}] line is
also very strong in NGC\,4039, while it has not been detected in
NGC\,4038.

In the mid-IR AGN activity can usually be traced by the high
excitation [\ion{O}{4}] and [\ion{Ne}{5}] emission lines
\citep{spi92,stu02,wee05}.  However, the best indicators, the
[\ion{Ne}{5}] lines at 14.3 and $24.3\mu$m, were not detected in
either nucleus.  The [\ion{O}{4}] line is not produced in measurable
quantities in \2~regions but can be excited by hot Wolf-Rayet stars,
shocks, and AGN.  In a study of starburst galaxies with {\sl ISO}
\citet{lut98} defined the ratio [\ion{O}{4}] /
([\ion{Ne}{2}]$+0.4\times$[\ion{Ne}{3}]), and found values ranging
between 0.006 (M82) and 0.062 (II\,Zw\,40) with a mean of 0.022.  As
can be seen from Table~\ref{tabfslines} our ratios for the peaks~1, 3,
4, 5 and 6 are 0.016, 0.026, 0.021, 0.040 and 0.021, respectively --
all in excellent agreement with the mean starburst value of
\citet{lut98}.  For the two nuclei of NGC\,4038 and NGC\,4039 we find
ratios of 0.030 and 0.035, respectively.  Both ratios are also well
within in the range covered by ``pure'' starburst systems and provide
no evidence for AGN activity.  This finding is further supported by
the large equivalent widths of the PAH features (Tab.~\ref{tabpahs}),
which would not be expected in the surroundings of an accreting black
hole \citep[e.g.,][]{wee05}.

\begin{figure}
\epsscale{1.1}
\plotone{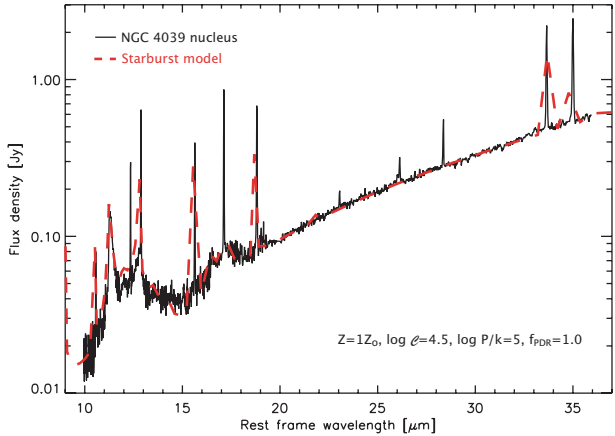}
\caption{{\sl IRS} spectrum of the NGC\,4039 nucleus (thin curve) with
          the best fitting starburst SED model template from
          \citet{gro08} (thick dashed curve).} 
          \label{fig4039nucmodel}
\end{figure}

Besides the high excitation lines, arguably the best mid-IR diagnostic
for the presence of an AGN is the spectral continuum.  Hot dust around
an AGN will significantly flatten the continuum slope (see e.g.,
Fig.~8 in \citet{bra06}).  In Fig.~\ref{fig4039nucmodel} we compare
the spectrum of the nucleus of NGC\,4039 to a starburst model of
\citet{gro08} chosen to best match the dust continuum and the PAH
features.  The determined parameters of the fit indicate that the
nuclear abundances are similar to Solar ($\sim 1$\Zsol ), with the
star formation occurring in somewhat distributed manner ($\log {\cal
  C}=4.5$) in regions with standard ISM pressures ($\sim \rm{P}_0 /
k=10^5$ K\,cm$^{-3}$), but still reasonably embedded within their
molecular gas birth clouds ($\sim f_{\rm PDR}\sim 1.0$).  (For a full
description of the parameters see \citet{gro08}).  The good match
between the observed continuum and a ``pure'' starburst model shows
that the spectrum of the nucleus of NGC\,4039 is fully consistent with
emission from dust that is heated by star formation only.


\subsection{The Properties of Individual SSCs}
\label{secpsscs}
Four of our six infrared peaks, namely peaks~1, 2, 3, and 5, are
located in the overlap region (Fig.~\ref{figslitpos}), which is the
region of the most intense star formation in the Antennae
\citep{mir98,vig96}; peaks~4 and 6 are located outside the overlap
region.  In this subsection we summarize briefly some of their
properties based on information from the literature.  Some cluster
masses are discussed in Section~\ref{secsfrs}.

Peak~1 coincides with a clump of molecular gas as traced by CO
\citep[peak CO\_S and SGMC 4--5, ][]{sta90,wil00,wil03}.  It is also
the brightest source at centimeter wavelengths, and the spectral index
of the radio emission indicates that the spectral energy distribution
is dominated by the thermal radiation from \2~regions \citep{hum86}
with an equivalent of $\sim$ 5000 O5~stars \citep{nef00}.  However,
the optical counterpart to this source corresponds to a very
inconspicuous, faint, red source in the HST images \citep[source
80]{whi95} obscured by large amounts of gas and dust.  Based on
sub-arcsecond near- and mid-IR spectroscopic data, both \citet{gil00}
and \citet{sni07} find an ionized gas density of $10^4$ cm$^{-3}$.

Peak~2 is the mid-IR counterpart to a bright and blue complex of
clusters, which contains eight optical sources within a region of
1\farcs5 \citep{whi95}, corresponding to 160~pc in projection at our
adopted distance.  Peak~2 also corresponds to the second brightest
radio source in the Antennae.  Like peak~1 it has a shallow spectral
slope, typical for the emission from \2~regions \citep{hum86,nef00}
with an equivalent of $\sim$ 3000 O5~stars.  The density of the
ionized gas has been estimated to be $\sim 10^4$~cm$^{-3}$
\citep{sni07}.

Peak~3 corresponds to the center of the largest concentration of
molecular gas in the overlap region. The coincident sources CO\_F and
part of SGMC~1 \citep{sta90,wil00,wil03} contain a total molecular gas
mass over $6\cdot 10^8$~\Msol.  The slope of the corresponding bright,
compact radio source \citep{nef00} is steep, arguing for a somewhat
older cluster with its radio emission being dominated by supernova
remnants.

Peak~4 is located to the far west of the nucleus of NGC\,4038.  The
radio observations show a moderately bright, compact source with a
steep spectral slope, indicative of an evolved stellar population
\citep{hum86,nef00}.

Peak~5 is located in the overlap region, approximately 1.6~kpc to the
north-west of peak~1, and is the brightest sub-millimeter peak.  It
coincides with the sources CO\_W and part of SGMC~1
\citep{sta90,wil00,wil03}, which are huge concentrations of
approximately $3.9\cdot 10^8$\Msol\ of molecular gas.  With an
A$_V\sim 11.81$ it is one of the most heavily embedded clusters in the
Antennae \citep{men05}.  Its optical counterpart is identified as the
red source 115 in \citet{whi95}.  Peak~5 also coincides with a bright,
compact radio source \citep{nef00}.  The spectral index suggests that
the dominant radio emitters are supernova remnants.

Peak~6, finally, is a mid-IR source located in a region of more
extended, fuzzy IR emission, to the east of the nucleus of NGC\,4038.
The spectral slope of the radio emission is at the division point
between the radiation field of \2~regions and supernova remnants.


\subsection{Cluster Ages}
The age estimates depend significantly on the method being used, and
show significant scatter.  The literature values on cluster ages of
all our clusters are summarized in Table~\ref{tabage}.  The various
age diagnostics indicate that the luminous IR sources in the northern
part of the overlap region are somewhat older than the ones in the
more active, southern overlap region.

\begin{deluxetable}{l r r r r r}
\centering 
\tabletypesize{\scriptsize} 
\tablecaption{Cluster Ages from the Literature}
\tablewidth{0pt} 
\tablehead{
\colhead{}&
\colhead{Age\tablenotemark{a}} &
\colhead{Age\tablenotemark{b}} &
\colhead{Age\tablenotemark{c}} &
\colhead{Age\tablenotemark{d}} &
\colhead{Age\tablenotemark{e}} \\
\colhead{Position} &
\colhead{[Myr]} &
\colhead{[Myr]} &
\colhead{[Myr]} &
\colhead{[Myr]} &
\colhead{[Myr]} 
}
\startdata 
peak 1   & 2.0     & $2.3 - 4.0$ & 3.5     & $\le 2.5$ & 2.5 \\
peak 2   & 3.8     & $2.3 - 4.0$ & \nodata & $\le 3$   & 3.0 \\
peak 3   & 2.0     & $3.2 - 4.9$ & 3.9     & $\le 3$   & 3.0 \\
peak 4   & 7.0     & $\ge 6$     & \nodata & \nodata   & 7.0 \\
peak 5   & \nodata & $3.7 - 5.1$ & 5.7     & $3-5$     & 4.5 \\
peak 6   & 8.4     & $4.3 - 5.7$ & \nodata & \nodata   & 6.0 \\
\enddata
\tablenotetext{a}{from H$_{\alpha}$ \citep{whi02}}
\tablenotetext{b}{from Br$_{\gamma}$ equivalent width \citep{men05}}
\tablenotetext{c}{from Br$_{\gamma}$ equivalent width \citep{gil07}}
\tablenotetext{d}{from the equivalent width of several hydrogen 
                  recombination lines \citep{sni07}}
\tablenotetext{e}{average age from the listed estimates}
\label{tabage}
\end{deluxetable}

\begin{figure}[ht]
\epsscale{1.1}
\plotone{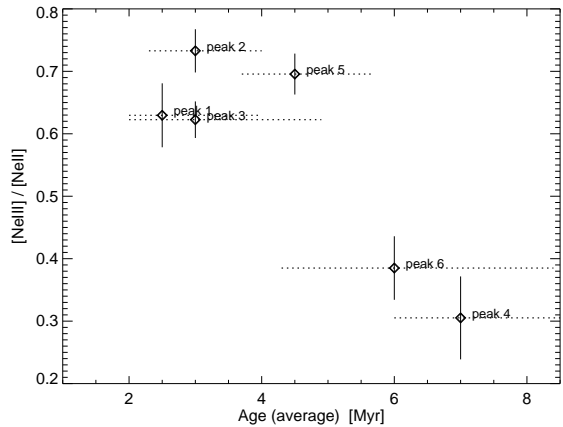}
\caption{The hardness of the radiation field (as measured by the line
         flux ratio of [\ion{Ne}{3}]\,/\,[\ion{Ne}{2}]) versus the
         average age of the SSCs from Table~\ref{tabage}.  The
         horizontal dotted lines indicate the age ranges covered by
         the various estimates. \label{figclusterage}}
\end{figure}

In Fig.~\ref{figclusterage} we compare the average cluster ages from
the literature (Tab.~\ref{tabage}) to the radiation hardness measured
by the [\ion{Ne}{3}]\,/\,[\ion{Ne}{2}] line ratio.  There is a clear
trend of older clusters displaying a softer radiation field due to the
disappearance of the early-type~O and Wolf-Rayet stars after about
5~Myr. While this result may not come as a surprise it is reassuring
that completely independent methods are consistent.

We note that starburst activity exists to both east (including peak~6)
and west (including peak~4) of the nucleus of NGC\,4038 at almost the
same level, in terms of the [\ion{Ne}{2}] and PAH intensity.
In fact, peaks~4 and 6 have almost identical mid-IR spectra
(Fig.~\ref{figspecall}), yet are separated by more than 6.6~kpc in
projection.  With a velocity dispersion of $\sim 10$~km\,s$^{-1}$
\citep{whi05} the dynamical time to connect the two points would be
647~Myr, about two orders of magnitude longer than the cluster ages.
This suggests that the dominant mode of triggering the formation of
SSCs does not propagate hydrodynamically but is governed by local
processes.


\subsection{Physical Conditions across the Antennae}
\label{secpropism}
In this subsection we discuss the spatial variations in the ISM as
revealed by the spectral maps shown in Fig.~\ref{figsixmaps}.  These
maps provide a more global picture of the physical conditions across
the central region of the Antennae galaxies.  Unfortunately, the
[\ion{Ne}{3}] line, which we utilize heavily in
Section~\ref{seccontslp}, is not covered by the {\sl IRS}-SL module
and would have to be observed through the much wider {\sl IRS}-LL
slit.  However, given that the excitation potential of [\ion{Ne}{3}]
is reasonably close to that of [\ion{S}{4}] (see
Table~\ref{tabfslines}), we substitute the [\ion{S}{4}] line as a
tracer of the harder radiation.

Fig.~\ref{figsixmaps} (top) compares the maps of [\ion{Ne}{2}] with
[\ion{S}{4}].  It takes only 21.6~eV to excite [\ion{Ne}{2}], and the
emission traces primarily the regions where star formation occurred in
the past $\sim 10$~Myr.  (As a rule of thumb we assume that
[\ion{Ne}{2}] is primarily associated with stars younger than $\sim
10$~Myr, while [\ion{S}{4}] traces stars younger than $\sim 4-6$~Myr
in a starburst).  The [\ion{S}{4}] emission originates predominantly
from two clusters peaks~1 and 2 in the overlap region, with peak~2
being even brighter in [\ion{S}{4}] than peak~1
(Fig.~\ref{figsixmaps}).  Although this has already been indicated by
the spectra in Fig.~\ref{figspecall}, and qualitatively known since
ISO \citep{mir98}, it is once again remarkable how the bulk of the
current massive star formation in an interacting system, which extends
over tens of kilo-parsecs, is confined to just two compact regions.  We
also detect some [\ion{S}{4}] emission from the nucleus of NGC4039,
but none from 4038 (Section~\ref{secnuclei}).  The [\ion{Ne}{2}] map
is also dominated by the two clusters peak~1 and 2, although to a much
lesser degree than the [\ion{S}{4}] map, and also shows emission from
the nucleus of NGC\,4038.  Noticeable [\ion{Ne}{2}] emission is also
detected to the west of the nucleus of NGC\,4038, where numerous older
SSCs reside \citep{men05}.

Overall, the PAH maps in Fig.~\ref{figsixmaps} (center row) resemble
the [\ion{Ne}{2}] map\footnote{See section~\ref{seclines} for a
  discussion of the contamination of the [\ion{Ne}{2}] map by
  $12.7\mu$m PAH emission.}, but show noticeably more diffuse
emission.  In ``normal'' galaxies a predominant fraction of the total
PAH emission arises from the diffuse ISM, heated only by the
interstellar radiation field, with a only a low fraction rising from
PDRs \citep{dra07b}. However in starbursting galaxies like the
Antennae the situation is reversed, with the major part of the total
emission coming from the regions of recent star-formation.  Here the
PAH maps trace primarily the PDRs and hence the environment of OB
clusters.  However, the fact that the [\ion{Ne}{2}] and PAH map
resemble each other qualitatively implies that we do not resolve the
\2~region/PDR interfaces at the spatial resolution of the {\sl IRS}-SL
spectral map, which is between 200 -- 380~pc, depending on wavelength
and orientation.

The most obvious difference between the [\ion{Ne}{2}] and the PAH maps
is the strength of peaks~1 and 2 relative to the nucleus of NGC\,4038.
The latter is the dominant source of PAH emission, while the embedded
SSCs play only a minor role.  Within the uncertainties, which are
dominated by the lower signal-to-noise in the $6.2\mu$m map, we find
no significant differences between the $11.3\mu$m and the $6.2\mu$m
PAH maps.  (We note that the PSF in the former is also 1.8 times
larger).  The similarity between the two PAH maps agrees with our
study of the PAH spectrum in Section~\ref{secpahvari}.

The bottom row in Fig.~\ref{figsixmaps} shows the spectral maps in the
H$_2$ S(2) and S(3) lines.  Both maps are very similar.  We find that
approximately 45\% of the total H$_{2}$\,S(3) emission comes from a
slightly extended region around the southern nucleus, NGC\,4039.  The
nucleus of NGC\,4038 does show some H$_2$ emission, but very little
compared to NGC\,4039.  The emission peaks of the clusters in the
overlap region are more compact and also weaker.  Most interesting
here is the small spatial offset of about two pixels, corresponding to
approximately 380~pc, that exists between the peaks of H$_2$ emission
and the spectral continuum, for both peak~5 and the nucleus of
NGC\,4039.  The H$_2$ S(2) emission peaks to the south in both objects
and appears also, at least near the nucleus on NGC\,4039, more
extended.  Since the offsets are with respect to the continuum
emission, which was simultaneously observed with the line emission,
they cannot be due to spacecraft pointing errors or other
observational effects.  Rather they must arise from a physical
displacement between the massive SSCs and the source of H$_2$
emission.

Finally, we compared the $8.6\mu$m PAH map with the {\sl IRAC} $8\mu$m
image of \citet{wan04}.  Despite the slightly better spatial
resolution of the {\sl IRAC} image both maps look very similar with
one exception: the emission from peak~1 is noticeably brighter in the
{\sl IRAC} image than in the $8.6\mu$m PAH map.  The spectra in
Fig.~\ref{figspecall} and the values in Table~\ref{tabpahs} show that
the PAH equivalent widths of peaks~1 and 2 are significantly below
those of the other clusters.  Although the {\sl IRAC} $8\mu$m band is
generally dominated by PAH emission, a proper continuum subtraction is
obviously important for quantitative results.


\subsection{Dust Temperature and Radiation Field}
\label{seccontslp}
\begin{figure}[ht]
\epsscale{1.1}
\plotone{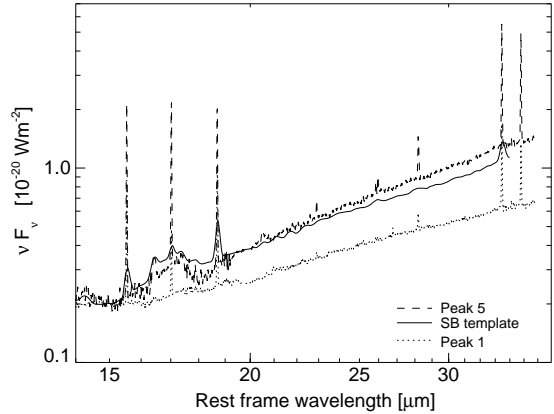}
\caption{Comparison of the $14 - 37\mu$m continuum slopes for three
         selected objects: peaks~1 and 5 with the most shallow and
         steep slopes in our sample of Antennae clusters, and the
         average starburst template spectrum from \citet{bra06}. The
         spectra are plotted in units of $\nu F_{\nu}$ to enhance the
         differences, and have been normalized to unity at
         $15\mu$m. \label{figcontcomp}}
\end{figure}

The mid-infrared continuum $\ge 20\mu$m is produced by the thermal
emission from dust heated by the stellar clusters.  The slope of the
continuum reflects the temperature distribution of the dust. The
presence of a hot dust component (100\,K $\le T \le$ 250\,K) flattens
the continuum slope in the wavelength range of $15 - 35\mu$m, while a
cold dust component ($T \le 50$\,K) steepens the slope.  The continuum
fluxes at $15\mu$m and $30\mu$m and their ratios are listed in
Table~\ref{tabspeccont}.  Peaks~1 and 2 have the shallowest slopes,
indicating the highest dust temperatures, while peaks~3 and 5 show the
steepest continua, evidence for a dominating dust component at lower
temperature.  The difference is also illustrated in
Fig.~\ref{figcontcomp} for the most extreme cases of peaks~1 and 5, in
comparison to the average starburst galaxy spectrum from
\citet{bra06}.  The slope of the starburst template lies between the
two Antennae clusters.

Generally, the dust temperature depends on a complex relation between
ISM density, grain size distribution, stellar properties, strength of
the radiation field, and the local geometry of cluster, \2~region,
PDR, and surrounding ISM.  We have investigated possible correlations
of the dust temperature with cluster properties from the literature
and our measurements, such as cluster age (Table~\ref{tabage}), mid-IR
luminosities (Table~\ref{tabspeccont}), extinction
(Table~\ref{tabext}), PAH strength (Table~\ref{tabpahs}), hardness of
the radiation field (Table~\ref{tabfslines}) or star formation rate
(Table~\ref{tablumis}).  However, no statistically convincing
correlation between these quantities has been be found.


\subsection{Cluster Luminosities and Star Formation Rates}
\label{secsfrs}
The mid-IR continuum fluxes can be used to estimate the total
infrared luminosity of starburst systems (in which $L_{\rm IR} \approx
L_{bol}$).  In their Section~4.2 \citet{bra06} discuss a method based
on the $15\mu$m and $30\mu$m continuum fluxes.  Adopting their
empirical relation and a distance of 22~Mpc \citep{sch08} we get
$L_{\rm IR} = 2.067\cdot F_{15} + 5.324\cdot F_{30}$ in units of
$10^9$\Lsol.  We note that this empirical relation was derived for
global starbursts, where the temperature components may different from
the one in and around more compact SSCs.  However, the good agreement
in Fig.~\ref{figcontcomp} between the slopes of the various systems
indicates that the estimate works reasonably well.

In Table~\ref{tablumis} we list the mid-IR derived cluster
luminosities.  The infrared luminosities of the eight compact regions
already add up to $L_{\rm IR} = 3.82\cdot 10^{10}$\Lsol.  The total
infrared luminosity of the Antennae, measured from the four {\sl IRAS}
bands at our distance, is $L_{\rm IR} = 7.2\cdot 10^{10}$\Lsol
\citep{san03}.  Hence, the eight individual regions studied here
account for slightly more than half of the total luminosity of the
system.

\begin{deluxetable*}{l c c c c c c}
\tabletypesize{\scriptsize} 
\tablecaption{Derived Cluster Properties}
\tablewidth{0pt} 
\tablehead{
\colhead{} &
\colhead{$L_{\rm IR}$\tablenotemark{a}} &
\colhead{SFR} &
\colhead{T(H$_{2}$)\tablenotemark{b}} &
\colhead{T(H$_{2}$)\tablenotemark{c}} &
\colhead{M(H$_{2}$)\tablenotemark{d}} &
\colhead{M(H$_{2}$)\tablenotemark{e}} \\
\colhead{Position} &
\colhead{[$10^9$\Lsol]} &
\colhead{[\Msol/yr]} &
\colhead{[K]} &
\colhead{[K]} &
\colhead{[$10^6$\Msol]} &
\colhead{[$10^6$\Msol]} 
}
\startdata 
nuc 4038 & 3.67 & 0.63 & 329       & 327 & 1.78    & 2.24 \\
nuc 4039 & 1.86 & 0.33 & 378       & 334 & 3.97    & 3.56 \\
peak 1   & 11.41 & 1.97 & \nodata   & 302 & \nodata & 3.09 \\
peak 2   & 10.40 & 1.81 & 288       & 267 & 2.82    & 3.20 \\
peak 3   & 4.35 & 0.74 & 311       & 300 & 1.97    & 3.39 \\
peak 4   & 1.29 & 0.22 & \nodata   & 374 & \nodata & 0.22 \\
peak 5   & 3.86 & 0.66 & 296       & 309 & 3.67    & 4.24 \\
peak 6   & 1.31 & 0.22 & \nodata   & 360 & \nodata & 0.30 \\ 
\enddata
\tablecomments{For uncertainties in the H$_2$ mass and temperature
               estimates see Section~\ref{secH2ana}. The S(3)flux 
               from peak~1 was well detected but is uncertain due 
               to a data artifact.}
\tablenotetext{a}{Luminosity estimate following \citet{bra06}}
\tablenotetext{b}{from the `lores' S(3) and S(2) line fluxes}
\tablenotetext{c}{from the `hires' S(2) and S(1) line fluxes}
\tablenotetext{d}{from the `lores' S(3) line fluxes and temperatures}
\tablenotetext{e}{from the `hires' S(2) line fluxes and temperatures}
\label{tablumis}
\end{deluxetable*}

\citet{ken98} has shown that the $8-1000\mu$m infrared luminosity
$L_{\rm IR}$ of starbursts is a good measure of the SFR as given by
\begin{displaymath}
\mbox{SFR}\ [M_{\odot}\mbox{yr}^{-1}] = 4.5\cdot
10^{-44} L_{\rm IR}\ [\mbox{erg\ s}^{-1}].
\end{displaymath}
We note that the SFR conversion given by \citet{ken98} strictly
applies only to dusty starbursts in the continuous star formation
approximation, with ages of order $10-100$~Myr.  Our clusters are
significantly younger than that, so in the continuous star formation
approximation (which is a reasonable one, given the more or less
continuous range of cluster ages, as seen in Fig.~\ref{figclusterage})
the IR luminosity per unit mass of stars formed will be somewhat
lower.  In other words, for regions younger than 10~Myr the
\citet{ken98} relation will systematically overestimate the SFRs.

Our estimated SFRs for the individual clusters are listed in
Table~\ref{tablumis}.  Co-adding the eight regions we get a total
$\mbox{SFR}\approx 6.6$\Msol\,yr$^{-1}$.  Given the compact sizes of
the regions our results agree with \citet{wan04}, who derived from
Spitzer-{\sl IRAC} $3-8\mu$m observations that the rates of star
formation in the active regions are as high as those seen in starburst
and some ultraluminous infrared galaxies on a ``per unit mass''basis.

As a cross-check we compare our SFRs to an independent estimate by
dividing the stellar masses of the clusters from the literature by
their ages.  For peak~1 \citet{men01} determined $3\cdot 10^6$\Msol\
and \citet{sni07b} found $1.1-1.2\cdot 10^6$\Msol.  For peak~2
\citet{men01} found $1.6\cdot 10^6$\Msol\ and \citet{sni07b} measured
$1.2-1.7\cdot 10^6$\Msol.  For average cluster masses of $2\cdot
10^6$\Msol\ and $1.5\cdot 10^6$\Msol\ for peaks~1 and 2, and ages of
2.5 and 3.0~Myr (Table~\ref{tabage}) we estimate (continuous) star
formation rates of 0.8 and 0.5\Msol\,yr$^{-1}$, respectively.  These
rates are roughly a factor of three below the SFRs derived from the
mid-IR fluxes.  However, we emphasize the substantial systematic
uncertainties in these estimates.  \citet{sni07b} assumed a Salpeter
IMF from $0.1 - 100$\Msol, while \citet{men01} assume a lower mass
cutoff of 1\Msol, corresponding to a 2.6 times smaller cluster mass.
Using the same IMF would lead to mass differences between those
studies of factors of 6.5 and 2.9 for peaks~1 and 2, respectively.  

The sum of the SFRs around the two nuclei and the six
infrared-brightest regions ($4.8\cdot 10^6$\,pc$^2$ in total) as
listed in table~\ref{tablumis} is 6.6\Msol\,yr$^{-1}$.  As stated
above, our method will likely provide an overestimate of the true SFR.
On the other hand, star formation will also occur outside our regions
at a reduced rate.  Hence, we consider our estimate of
6.6\Msol\,yr$^{-1}$, within 50\%, as representative for the
SFR in the Antennae galaxies.

We also note that \citet{zha01} determined a rather large SFR of
26\Msol\,yr$^{-1}$ for a distance of 22~Mpc.  Using the \citet{ken98}
relation between SFR and $L_{\rm IR}$ this would correspond to a more than
two times higher total infrared luminosity than the one derived from
the {\sl IRAS} bands \citep{san03}, and would move the Antennae to the
class of LIRGs.  Hence, we consider the SFR derived by \citet{zha01}
an overestimate.

Recently, \citet{smi07c} investigated Spitzer photometry of a sample
of 35 tidally distorted pre-merger interacting galaxy pairs selected
from the Arp atlas.  These authors found a very modest enhancement of
the SFRs at early stages of $\sim 1$\Msol~yr$^{-1}$ on average.
Although the Antennae is still in an early stage of interaction
\citep{mih96} it shows an enhanced SFR, well above the sum of typical
star formation in two normal spiral galaxies.  With its ample amount
of molecular gas \citep{gao01} the system is likely to become an even
more spectacular starburst after one or two orbits.


\subsection{PAH Strength and Aperture Effects}
\label{secpahstudies}
PAHs are considered the most efficient species for stochastic,
photoelectric heating by UV photons in PDRs \citep{bak94}.  They are
usually a good tracer of starburst activity in a statistical sense
\citep[e.g.,][ and references therein]{bra06}.  However, numerous
studies over the past two decades \citep[e.g.][]{geb89,ces96,tra98}
have shown that intense UV fields can lead to the gradual destruction
of PAH molecules.  A clear anti-correlation between the $11.3\mu$m PAH
strength and the radiation field across the central region of NGC~5253
was recently observed by \citet{bei06}.  We investigate this effect
for the Antennae in two different ways: via a pixel-by-pixel analysis
within the spectral map and for the eight `hires' spectra.

\begin{figure}[ht]
\epsscale{1.1}
\plotone{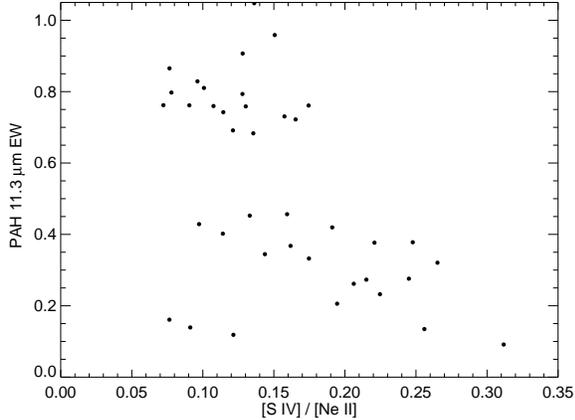}
\caption{The relative strength of the $11.3\mu$m PAH feature
         (a normalized ``line''-to-continuum ratio) versus the hardness
         of the radiation field as expressed by the line ratio of
         [\ion{S}{4}]\,/\,[\ion{Ne}{2}].  Each data point corresponds
         to one pixel in the spectral map.  Only those 42 pixels with
         a signal of greater than four standard deviations in the
         \ion{S}{4} map have been included.  \label{figmaphardnessew}}
\end{figure}

The former is shown in Fig.~\ref{figmaphardnessew} where the
$11.3\mu$m PAH-to-continuum ratio is plotted versus the hardness of
the radiation field as expressed by the line ratio of
[\ion{S}{4}]\,/\,[\ion{Ne}{2}].  Each data point in
Fig.~\ref{figmaphardnessew} corresponds to a spatial pixel in our
spectral map for which the signal-to-noise at [\ion{S}{4}] is greater
than four standard deviations.  (See section~\ref{seclines} concerning
the contamination of [\ion{Ne}{2}] by the $12.7\mu$m PAH emission in
`lores' mode).  Fig.~\ref{figmaphardnessew} reveals a trend that
regions with harder radiation fields show relatively weaker PAH
emission.

We note that the strength of the [\ion{S}{4}] relative to the
[\ion{Ne}{2}] line does not only depend on the hardness of the
radiation field, but is sensitive to the gas density as well. Given
the critical densities of $3.7\cdot 10^4$~cm$^{-3}$ and $6.1\cdot
10^5$~cm$^{-3}$ for [\ion{S}{4}] and [\ion{Ne}{2}], respectively
\citep[][and references therein]{tie05}, it is expected that
densities above $10^4$~cm$^{-3}$ will decrease the
[\ion{S}{4}]\,/\,[\ion{Ne}{2}] ratio.  Model calculations with {\sl
CLOUDY} show that an increase in density from $10^4$ to
$10^5$~cm$^{-3}$ can cause a decrease in
[\ion{S}{4}]\,/\,[\ion{Ne}{2}] by a factor of three, contributing
significantly to the scatter in Fig.~\ref{figmaphardnessew}.

To quantify the strength of the radiation field we use the line fluxes
of [\ion{Ne}{2}] and [\ion{Ne}{3}] and the parametrization of
\citet{bei06}:
\begin{displaymath}
\left( F_{\rm [Ne\,II]12.8\mu m}\,+\,F_{\rm [Ne\,III]15.6\mu m}\right)
\times
\frac{F_{\rm [Ne\,III]15.6\mu m}}{F_{\rm [Ne\,II]12.8\mu m}}.
\end{displaymath}
The first factor, $F_{\rm [Ne\,II]12.8\mu m}\,+\,F_{\rm
  [Ne\,III]15.6\mu m}$, is a measure of the intensity of the field,
assuming that all Neon exists in one of the two ionization states.
The second factor, $\frac{F_{\rm [Ne\,III]15.6\mu m}}{F_{\rm
    [Ne\,II]12.8\mu m}}$, is a measure of the radiation field
hardness.  We define the product of intensity and hardness as the
``strength'' of the radiation field.

\begin{figure}[ht]
\epsscale{1.1}
\plotone{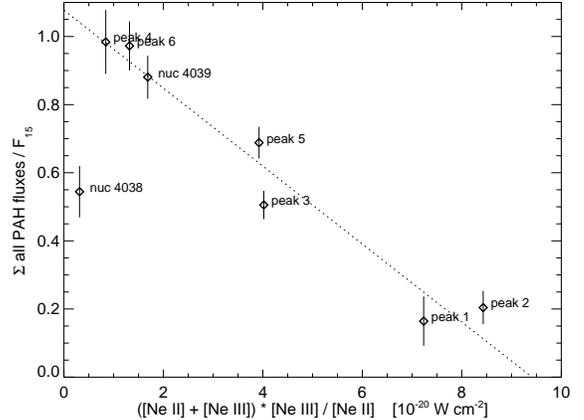}
\caption{The relative strength of all PAH features within the $10 -
         20\mu$m range versus the strength of the radiation field.
         \label{figpahradstrength}}
\end{figure}

In Fig.~\ref{figpahradstrength} we plot the strength of the PAH
features, measured from the `hires' spectra, against the strength of
the radiation field.  On the ordinate we plot the average PAH strength
calculated by co-adding all the PAH fluxes listed in
Table~\ref{tabpahs} and normalizing them to the $15\mu$m continuum
flux (Table~\ref{tabspeccont}).  There is a clear linear
anti-correlation between the strength of the PAH emission and the
radiation field.  The only outlier is the inactive nucleus of
NGC\,4038.

Both rather independent methods and data sets provide statistical
evidence that the strength and hardness of the radiation field affect
the PAH emission.  However, we cannot be certain that this effect is
actually due to a destruction of PAH molecules in stronger radiation
fields.  The sources with the strongest PAH ``equivalent width'' are
peaks~4 and 6.  Both are located in less dense regions.  These
clusters are also slightly older and have had more time to shape the
surrounding ISM via stellar winds and SNe.  If the size of the
emitting region is radiation-bounded these regions may just have
``grown'' more complex \2~region/PDR interfaces from where the PAH
emission originates. In other words, the PDR surface area is higher
relative to the thermal emission from dust in the denser regions,
which produces the underlying continuum.

Unfortunately, with the wide {\sl IRS}-SH slit width, corresponding to
500\,pc, we cannot resolve the \2~region/PDR interfaces.  However,
some of the clusters were recently observed by \citet{sni06} with {\sl
  VISIR} on ESO's VLT using a sub-arcsecond slit corresponding to only
75\,pc at our distance of 22~Mpc.  Comparing our {\sl IRS}-SH spectra
of peaks~1 and 2 to their {\sl VLT-VISIR} spectra, \citet{sni06} found
that, while the continuum fluxes observed with both instruments are
quite similar, the wider {\sl IRS} slit detects much stronger PAH
emission.  The individual clusters are by far the most luminous
sources on projected scales of a few hundred parsecs, and we may
observe the profound effect of the OB clusters on their surrounding
ISM.  However, \citet{sni06} concluded that a large fraction of the
PAH emission cannot be directly associated with the SSCs but
originates from an extended region around the cluster.  In fact, if
the space density of young clusters near peaks~1 and 2 is
significantly higher than elsewhere it could be that the extended PAH
emission, which is only picked up by the {\sl IRS}, is unrelated to
the main clusters and externally excited by smaller clusters in the
immediate vicinity of peaks~1 and 2.  At this point we cannot
distinguish between a giant \2~region of a hundred or more parsecs in
diameter and a locally enhanced density of OB clusters.  A significant
improvement will have to wait for the next generation of extremely
large telescopes (ELTs).


\subsection{(In-)Variability of the PAH Spectrum}
\label{secpahvari}
We have also investigated possible variations of the PAH spectrum.
While -- to first order -- the strengths of the PAH features are
observed to scale with each other \citep[e.g.,][Fig.~6]{smi07b}, the
PAH spectrum is expected to depend on the grain size distribution
\citep{dra01,dra07a}: smaller PAHs are more likely to emit at shorter
wavelengths while larger PAHs are more likely to produce stronger
features at longer wavelengths.  We utilize the longest baseline
provided by the `hires' spectra between strong PAH features and
selected the $11.3\mu$m feature and the $17\mu$m PAH
complex\footnote{The $17\mu$m PAH complex includes the $16.4\mu$m
feature \citep{mou00}, the $17.4\mu$m feature \citep{stu00}, and a
very broad feature at $17.1\mu$m, which contains over 80\% of the PAH
flux of that complex \citep{smi04}.}.  We find an average
$17\mu$m\,/\,$11.3\mu$m PAH ratio of 0.4 (Table~\ref{tabpahs}) with
some scatter but no significant correlation with any other observed
quantity.  From the {\sl SINGS} sample of galaxies \citet{smi07b} have
found that the $17\mu$m\,/\,$11.3\mu$m PAH ratio increases with
increasing metallicity.  Assuming solar abundance for the Antennae
\citep{bas06} our PAH ratio of 0.4 is consistent with the measurements
of \citet[][Fig.~16]{smi07b} although somewhat on the low side.

Another cause of potential variations of the PAH spectrum may be
ionization effects.  Several authors
\citep[e.g.,][]{ver96,ver02,for03} have reported on variations in the
PAH spectrum, namely that the C--C stretching modes at $6.2\mu$m and
$7.7\mu$m are stronger in ionized PAHs, relative to the bending mode
at $11.3\mu$m -- by factors of up to two.  Here we use the ratio of
the $8.6\mu$m C--H in-plane bending mode to the $11.3\mu$m C--H
out-of-plane bending mode, which is also sensitive to the charge state
of the PAH molecule \citep{hud95,job96}.  We have chosen these two for
a direct comparison because of their proximity in wavelength (same
beam size), similar susceptibility to dust extinction, and because
both are observed through the same {\sl IRS} (sub-)slit.
\citet{smi07b} investigated the $7.7\mu$m\,/\,$11.3\mu$m PAH ratio as
a function of the hardness of the radiation field and found no
significant correlation for the \2-nuclei in the {\sl SINGS} galaxies
sample (their Fig.~14).

\begin{figure}[tb]
\plotone{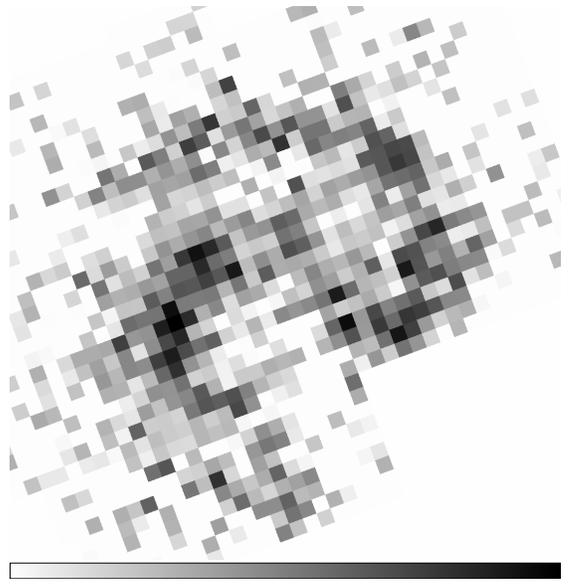}
\caption{{\sl IRS} spectral map of the $8.6\mu$m\,/\,$11.3\mu$m PAH
         ratio.  The median ratio has been normalized to unity, which
         is colored in mean gray, and the range extends over a factor
         of two to both sides.  Brighter regions represent relatively
         stronger $8.6\mu$m emission, darker regions weaker $8.6\mu$m
         emission.  Field size and orientation are as in
         Fig.~\ref{figsixmaps} but the resolution has been reduced by
         rebinning by a factor of two to suppress the noise.
         \label{figmappahratios}}
\end{figure}

Fig.~\ref{figsixmaps} already indicates that, to first order, both PAH
maps trace each other quite well.  Thus we have computed the ratio on
a pixel-by-pixel basis between the two $8.6\mu$m\,/\,$11.3\mu$m PAH
maps.  The map is shown in Fig.~\ref{figmappahratios} at two times
lower resolution to reduce the noise.  Regions with relatively
stronger $8.6\mu$m emission appear brighter, regions with relatively
stronger $11.3\mu$m emission appear darker in this ratio map.  While
the emission from the nucleus of NGC\,4038 appears in average ``gray''
the peaks~3, 5, and 6 appear darker and have values about 20\% above
the mean.  On the other hand, peaks~1 and 2, the most massive
clusters, appear brighter by about the same amount.  These variations
are not correlated with extinction effects, which would affect the
values in the opposite direction.  However, we emphasize that the
noise, in particular in the $8.6\mu$m map (Fig.~\ref{figsixmaps}), is
quite significant.  Although these variations may provide weak
evidence for changes in the PAH spectrum on cluster scales, a
conclusive analysis requires better resolution and lower noise.

\subsection{Strength of the H$_2$ Emission}
\label{secH2strength}
Molecular hydrogen is the most abundant molecule in the Universe.  It
plays a central role in star formation, not only as the major
ingredient to build up a star but also as the coolant to permit an
isothermal collapse of the gas cloud.  The density and temperature of
the H$_{2}$ molecule are of utmost importance for the processes in
starbursts.  However, the H$_{2}$ molecule is symmetric, has no dipole
moment, and the mid-IR lines originate from quadrupolar rotational
transitions.  Consequently, the lines are intrinsically weak and
usually hard to detect.  Since the first study of the S(2) line in
Orion by \citet{bec79} and the detection of the S(0) line in NGC~6946
by \citet{val96}, the mid-IR rotational lines in starbursts have been
studied by numerous groups.  Most relevant in our context are the {\sl
ISO-SWS} studies by \citet{rig02}, the review by \citet{hab05}, and
the Spitzer-{\sl IRS} studies of ULIRGs by \citet{hig06}, and {\sl
SINGS} galaxies by \citet{rou07}.

The H$_{2}$ lines in the Antennae are amongst the strongest detected
lines (Fig.~\ref{figspecall}).  We have calculated the total,
spatially integrated H$_{2}$\ S(2) and S(3) line fluxes by summing up
all pixels where the signal-to-noise was greater than three standard
deviations in the spectral maps and compared them to the sum of the
regions listed in Table~\ref{tabh2lines}.  In the S(3) line
essentially all of the measured flux of $6.1 \cdot 10^{-20}$ W
cm$^{-2}$ originates from our designated regions.  The emission in the
S(2) line is slightly more extended and the spatially integrated
measure picks up 27\% more flux than listed in Table~\ref{tablumis}.
We conclude that most of the total amount of warm H$_2$ over the
central Antennae system is concentrated around our eight sources.

\begin{figure}[tb]
\plotone{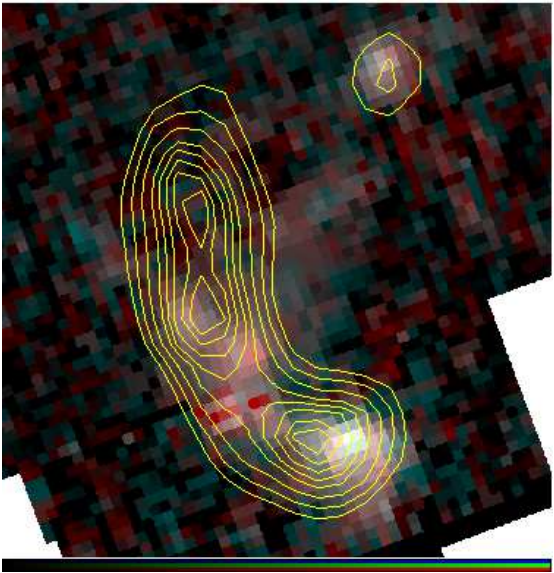}
\caption{False-color {\sl IRS} spectral map of the H$_2$ S(3) line in
         the green and blue channel, and the S(2) line in the red
         channel in square-root scaling.  Bluer colors mean warmer
         H$_2$.  North is up. The two nuclei and the emission from the
         clusters in the overlap region are clearly visible.  The
         yellow contour lines indicate the {\sl ISOCAM-CVF}
         measurements of the H$_2$~S(3) line by \citet{haa05}; the
         levels are at 39\%, 49\%, 58\%, 68\%, 78\%, 84\%, 90\%, and
         96\% of the maximum intensity (see Fig.~2 in
         \citet{haa05}).\label{figH2haas}}
\end{figure}

Fig.~\ref{figH2haas} shows a false-color image of the {\sl IRS} maps
with the H$_2$ S(3) and S(2) lines in blue/green and red.  Both nuclei
and the emission from the active overlap region are clearly visible.
The map shows that the largest portion (approximately 45\%) of the
total H$_2$\,S(3) emission comes from a region around the southern
nucleus.  The strongest H$_2$ emitter in the active overlap region is
peak~5.

We note that our integrated H$_{2}$\ S(3) line flux of $6.1 \cdot
10^{-20}$ W\,cm$^{-2}$ is about five times below the S(3) line flux of
$3.3 \cdot 10^{-19}$ W\,cm$^{-2}$ which \citet{haa05} have found.  In
Fig.~\ref{figH2haas} we overlaid the contours of their {\sl ISO}
H$_2$\,S(3) intensity map for comparison.  While our peak~5 is close
to their southern peak of the contour map, the northern peak of the
contour map has no luminous counterpart in our map.  Our closest peak
is peak~6, which contains more than an order of magnitude less H$_2$
mass than peak~5.  We have also checked the {\sl IRS}-LL map, which
includes the H$_{2}$S(1) emission line, albeit at two times lower
spatial resolution, and confirmed the strong drop in H$_2$ emission
northward of peak~3.  Altogether, we cannot confirm the claim by
\citet{haa05} of strong H$_2$ emission from a region extending farther
to the north of the overlap region, well beyond the most active area.
These differences in distribution and total line fluxes have
significant impact on the derived H$_2$ masses for the Antennae
system.

At the distance of 22~Mpc, our S(3) line flux corresponds to a line
luminosity of $9.2 \cdot 10^6$\Lsol.  Using the far-IR luminosity
derived by \citet{san03} we calculate the ratio to be
$L(\mbox{H}_{2})/L_{\rm FIR}\approx 1.6\cdot 10^{-4}$.  This
value is only about two times higher than the one for M82 and not
atypical for luminous starbursts and ULIRGs \citep[cf.][
Fig. 3]{haa05}.

Here we also investigate with what other observables the strong H$_2$
emission might be correlated.  \citet{rig02} found a good correlation
between the $7.7\mu$m PAH emission and the H$_2$\,S(1) line luminosity
in ULIRGs.  A tight correlation between H$_2$ and PAH emission was
also reported by \citet{rou07} for the {\sl SINGS} galaxies (their
Figs.~8c and 9).  In fact, they found that ``among the measured dust
and gas observables, PAH emission provides the tightest correlation
with H$_2$''.  This sounds plausible since both PAHs and H$_2$ reside
preferentially at the outer skin of PDRs, and are both predominantly
exited by FUV photons.  However, while PAH emission is almost
exclusively excited by UV photons, H$_2$ can be excited by shocks or
collisionally excited if the critical density of typically $\sim
10^3$\,cm$^{-3}$ is exceeded (Section~\ref{secH2excite}).  This may
well be given locally for the embedded SSCs where electron densities
up to $4\cdot 10^4$\,cm$^{-3}$ \citep{sch07} or even higher
\citep{mir98} have been measured.

\begin{figure}[tb]
\epsscale{1.1}
\plotone{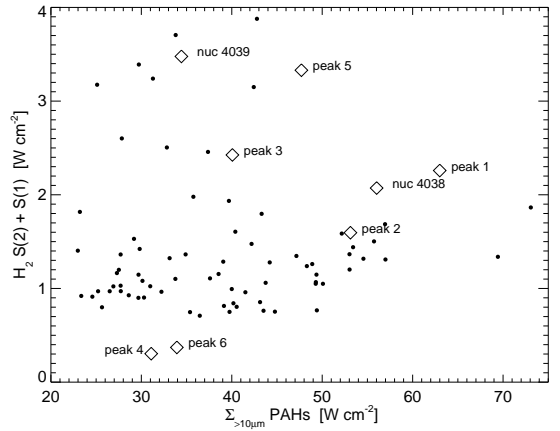}
\caption{The strength of the H$_2$ lines plotted versus the strength
         of the PAH emission features.  The figure contains two sets
         of data: the large diamonds correspond to the eight `hires'
         positions for which all PAH features longward of $10\mu$m
         have been added; the dots refer to pixels within our `lores'
         spectral maps for which the signal is at least $2.5\sigma$.
         In the case of the `lores' data the PAH strength is the sum
         of the $8.6\mu$m and $11.3\mu$m maps and the H$_2$ strength
         has been derived from the S(2) and S(3) lines.  The pixel
         fluxes have been scaled to match the `hires' data points.
         \label{figmaph2pah}}
\end{figure}

In Fig.~\ref{figmaph2pah} we investigate a possible correlation
between the H$_2$ line strength and the PAH strength.  The comparison
has been done for the eight sources observed in `hires' mode and for
the spectral maps on a pixel-by-pixel basis.  For both cases we do
{\sl not} find a tight correlation.  In fact, the apparently
uncorrelated scatter in Fig.~\ref{figmaph2pah} is remarkable, although
not completely unexpected given the different morphology between the
PAH and H$_2$ maps in Fig.~\ref{figsixmaps}.  We note that
\citet{rou07} base their PAH index on the $7.7\mu$m feature while we
use different PAH features.  However, since we do not observe any
significant variations within the PAH spectrum
(section~\ref{secpahvari}) we assume that both measures are
equivalent.

There are two possible explanations for the difference between the
previously observed PAH-H$_2$ correlation and our results.  First, the
sources of strong H$_2$ emission in the Antennae are compact regions
of a few hundred parsecs at most.  Larger scatter around the average
galactic value is therefore not unexpected \citep{rou07}.  Second, we
cannot rule out contributions from local shocks caused by supernovae.
The latter scenario will be discussed in Section~\ref{secH2excite}.


\subsection{H$_2$ Temperatures}
\label{secH2temp}
From a sample of 77 ultra-luminous infrared galaxies (ULIRGs) observed
with the {\sl IRS} \citet{hig06} measured a mean temperature of the
warm H$_2$ gas of 336\,K.  For the more quiescent galaxies in the {\sl
  SINGS} sample, \citet{rou07} determined a median temperature of only
154\,K.  However, the slit apertures of the {\sl SINGS} observations
sample the circum-nuclear regions; often these spectra have
contributions from multiple emitting sources, and presumably regions
with a range of densities and temperatures.  For the Antennae galaxies
\citet{kun96} derived a temperature of 405\,K from {\sl ISO-SWS} data,
whereas \citet{fis96} used {\sl ISO-LWS} (while using different
tracers) and derived a temperature for the PDR of $200\pm 60$\,K.  We
note that the common decomposition in warm and cold H$_2$ components
may not be unique and the absolute temperatures not necessarily
physically distinct.  However, they allow useful qualitative
comparisons between different systems.

Our temperature estimates for the different regions are listed in
Table~\ref{tablumis}.  We emphasize that the two independent
temperature estimates for the IR peaks which were derived from two
different sets of lines agree remarkably well, to within a few
percent.  This is likely because the individual spectra are dominated
by single sources, which almost completely thermalize the warm H$_2$.
Such a good agreement between temperature estimates from the S(1),
S(2) and S(3) lines would not be expected if the H$_2$ were primarily
shock-excited \citep{app06}.

To compare our estimates with the values from the literature we
``simulated'' a larger slit aperture (e.g., of {\sl ISO-SWS}) by
calculating a luminosity-weighted mean temperature from the S(2) and
S(1) lines for the clusters peak~1, 2, 3, 5, and 6 within the overlap
region.  This yields a mean warm H$_2$ temperature of $302\pm 26$\,K,
which lies significantly below the temperature estimate of
\citet{kun96} for the Antennae, but only 10\% below the average
temperature of ULIRGs \citep{hig06}.  The quoted uncertainty on the
temperature is the $1-\sigma$ standard deviation between the temperature
estimates in Table~\ref{tablumis}, and includes the temperature
variations between clusters as well as the independent measurements
from different {\sl IRS} modules.  Hence, we consider this uncertainty
as a good estimate of the total systematic uncertainties.

\begin{figure}[tb]
\epsscale{1.1}
\plotone{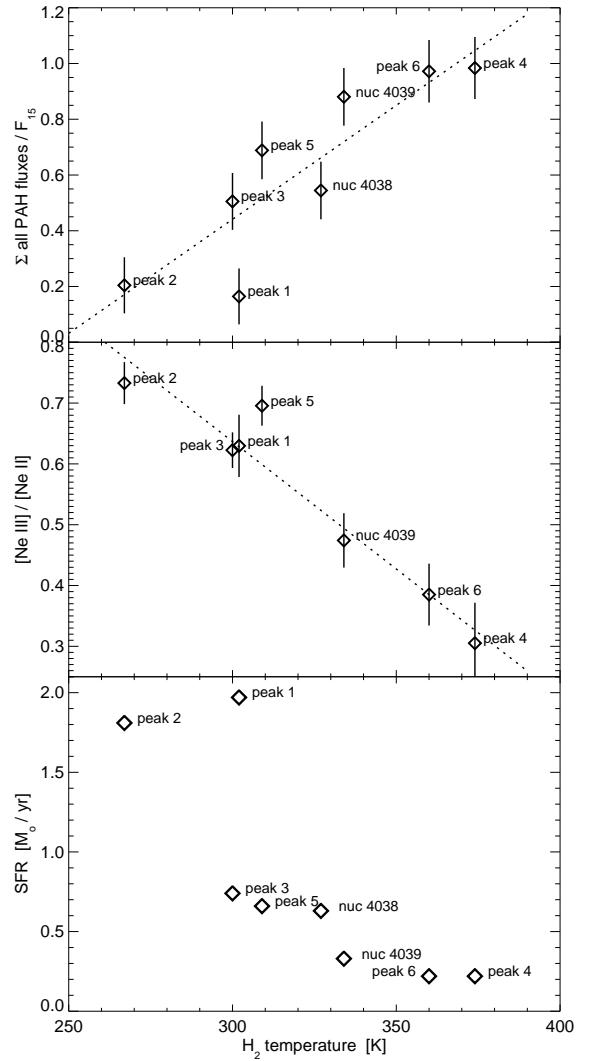}
\caption{Dependency of three important parameters on the temperature
         of the H$_{2}$ gas, calculated from the S(2) and S(1) lines.
	 {\sl Top:} The relative strength of the PAH emission (given
         by the flux sum of all PAH features between $10 - 20\mu$m
         normalized to the continuum flux at $15\mu$m).
	 {\sl Center:} The hardness of the radiation field (as given
         by the line flux ratio of [\ion{Ne}{3}]\,/\,[\ion{Ne}{2}]).
         One outlier, the nucleus of NGC\,4038, which shows a very low
         [\ion{Ne}{3}]\,/\,[\ion{Ne}{2}] ratio, is not included.  
	 {\sl Bottom:} The star formation rate as listed in 
	 Table~\ref{tablumis}.  The dotted lines are linear fits but 
	 only intended to guide the eye.}
	 \label{figH2threeplots}
\end{figure}

We also investigated the possible dependency of the temperature of the
warm H$_2$ component on other observables.  Fig.~\ref{figH2threeplots}
shows the relative strength of the PAH emission, the hardness of the
radiation field, and the star formation rate as a function of the
H$_2$ temperature.  All three plots show interesting correlations.

For each of the eight regions we have computed the sum of all PAH
features between $10 - 20\mu$m and normalized them to the continuum
flux at $15\mu$m.  The relative PAH strength correlates well with
the H$_2$ temperature.  It appears likely that those PDRs, which are
stronger PAH emitters relative to the continuum emission, have
warmer molecular gas.  On the other hand, the hardness of the
radiation field (center plot in Fig.~\ref{figH2threeplots}), expressed
in terms of [\ion{Ne}{3}]\,/\,[\ion{Ne}{2}], anti-correlates with the
H$_2$ temperature: regions of higher temperature show a softer
radiation field.  The general decrease toward hotter H$_2$ is also
observed individually for both line fluxes of [\ion{Ne}{2}] and
[\ion{Ne}{3}], but is steeper for the latter, resulting in the
decrease of radiation hardness in hotter regions.

These findings may, at first, be surprising as one might have guessed
that a harder radiation field is more energetic and will thus produce
a hotter environment.  Our favored explanation is that the older, more
evolved clusters had more time to shape the surrounding ISM via
stellar winds and supernovae, and the surrounding, clumpy dust and gas
clouds have evolved into larger entities.  The UV photons from the
cluster can penetrate farther into the cloud and provide a more
uniform temperature distribution over a larger area.  Very young
clusters may have hotter gas and dust near their inner shell but are
colder at their well shielded outskirts.  This is mainly a geometrical
argument, similar to the scenario discussed in
Section~\ref{secpahstudies}.  We also note that the top plot shows the
{\sl relative} PAH strength, normalized by the continuum flux at
$15\mu$m.  The absolute PAH fluxes, however, show a similar trend as
the Neon lines, namely stronger emission from regions of colder gas.
This indicates that the hotter regions are more efficient PAH
emitters.

Finally, the bottom plot in Fig.~\ref{figH2threeplots} shows the star
formation rate (Table~\ref{tablumis}) versus the H$_2$ temperature.
Although less significant than in the upper two plots there is a clear
trend of lower SFR for hotter regions.  This is consistent with the
previous anti-correlation between both hardness and line fluxes --
which are expected to be stronger in regions of recent star formation,
and the H$_2$ temperature.  Since the SFRs have been derived
independently the good agreement with our line diagnostics underlines
the reliability of the results.


\subsection{H$_2$ Masses}
\label{secH2mass}
The fraction of the warm to total molecular hydrogen mass is of
particular interest as it provides information on the dominant
excitation mechanism and can be compared to other galaxies.  The total
molecular hydrogen mass is dominated by the cold component, which is
derived from $^{12}$CO measurements and then converted to H$_2$ using
the standard Galactic conversion factor.  

For starburst-dominated galaxies \citet{rig02} found that the warm
H$_2$ ($T\sim 150$\,K) accounts for about $1 - 10$\% of the total
molecular hydrogen mass.  In Seyfert galaxies this fraction can be
higher \citep[$2 - 35$\%][]{rig02}.  Recently, \citet{sch07} reported
that the gas properties in the Antennae do not deviate significantly
from the $N_{H_2}/I_{CO}$ ratio found for the disk of the Milky Way,
and the conversion from CO is thus reasonably accurate.  \citet{wil00}
used a CO to H$_{2}$ conversion factor of $3 \cdot 10^{20}$
H$_{2}$\,cm$^{-2}$, and derived a total H$_{2}$ mass of $7.1 \cdot
10^9$\Msol, adopted to our distance of 22~Mpc.  Earlier estimates by
\citet{sta90} yielded a total H$_2$ mass of $1.3\cdot 10^9$\Msol.

The derived masses of warm H$_2$ are listed in Table~\ref{tablumis}.
We note that the two independent estimates from different {\sl IRS}
modules, H$_2$ lines and observing modes agree within 15\%.  Adding up
the warm molecular gas masses for the `hires' positions in
Table~\ref{tablumis} yields $M_{H_2}^{regions} = 2.0\cdot 10^7$\Msol.
If we also account for the more extended ``missed'' gas (approximately
25\% in the S(2) line -- Section~\ref{secH2strength}) we get
$M_{H_2}^{total} = 2.5\cdot 10^7$\Msol.  Based on the total mass
estimate from \citet{wil00} we measure a fraction of warm-to-total
molecular gas mass in the Antennae of 0.35\%.  This ratio needs to be
compared to the extreme case of NGC\,6240 where \citet{vdw93,arm06}
found about 15\% of the total molecular gas in its warm phase.
However, the Antennae appears to have, for its luminosity, an average
fraction of warm-to-total molecular gas mass.

These estimates can also be made for individual regions.  The
supergiant molecular clouds labelled SGMC~1, SGMC~2 and (SGMC~3 +
SGMC~4 + SGMC~5) by \citet{wil00} correspond approximately to our
peak~3, peak~5, and (peak~1 + peak~2), respectively.  Moving both
systems to the same distance, we calculate fractions of warm to total
gas masses of 0.4\%, 0.8\%, and 0.6\% for the three regions,
respectively.  The more localized ratios are somewhat higher than the
galactic average, which is not unexpected since the cold
H$_2$ is likely to cover a larger volume.  Similar results are
obtained from a comparison with the regions `south clump', `west
clump', and `east clump' in \citet{sta90}, which correspond
approximately to our (peak~1 + peak~2), peak~5, and peak~3,
respectively.


\subsection{H$_2$ Excitation Mechanisms}
\label{secH2excite}
All the observable lines at mid-IR wavelengths originate from
quadrupolar rotational transitions.  The rotational levels of H$_{2}$
can be populated in various ways: H$_2$ molecules may form in an
excited state \citep{tak01}; X-ray photons from energetic sources like
AGN or supernova remnants can ionize and heat the gas, which excites
the H$_2$ molecules via collisions with electrons and/or hydrogen
atoms \citep{lep83,dra92}; FUV photons can pump the H$_2$ molecules in
excited states \citep{bla87,hol97}; and excitation via shocks in the
ISM \citep{shu78,dra83}.  In starburst environments we can neglect the
former two scenarios and concentrate on UV pumping and shock
excitation.

UV pumping requires photons with energies $E_{\gamma}=6-13.6$\,eV to
excite the H$_2$ molecules, which cascade via vibration-rotational
transitions, producing characteristic fluorescent spectra.  Shocks are
expected to occur on various scales.  On small scales, shocks from
stellar outflows and supernova remnants can excite H$_{2}$
\citep{shu78}.  On larger, galactic scales streaming motions and
cloud-cloud collisions, in particular in merging systems have been
proposed \citep{vdw93,gil00,haa05,arm06,app06}.

Both FUV pumping and shock excitation have been shown to play a
dominant role in the excitation of H$_{2}$ in starburst environments
\citep[e.g.,][and references therein]{moo94,rou07}.  Pure fluorescent
spectra can usually be distinguished from thermalized, shock heated
spectra from the relative line strengths.  However, depending on the
critical densities, collisional de-excitation can also become
important for UV pumped H$_{2}$ states, leading to thermalized lower
levels.  Most of the emission of warm H$_{2}$ is likely to come from
dense PDRs with densities $\ge 10^3$\,cm$^{-3}$, and we expect that
the observed S(0) through S(3) lines are thermalized \citep{bur92}.
The near-IR rotation-vibration transitions 2-1 S(1) at $2.248\mu$m and
1-0 S(1) at $2.122\mu$m can be used to discriminate between shocks and
UV excitation \citep[e.g.,][]{tak00}.  Their ratio is typically 0.1 in
shocked regions (with the gas in LTE at $T\sim 2\cdot 10^3$~K), but
$\sim 0.6$ for ``pure'' UV fluorescence \citep{bla87}.  However, in
very dense PDRs ($\sim 10^5$ cm$^{-3}$) even the near-IR
rotation-vibrational lines are not a good discriminant anymore as
thermal collisions can transfer the lower-level ($\nu < 2$)
populations toward that in LTE, and will resemble the ratio observed
in shocked regions \citep[e.g.,][]{ste89}.

The observational evidence in the Antennae so far is confusing.
\citet{sni07b} derived 2-1\,S(1) / 1-0\,S(1) values between 0.15 and
0.34 for peaks~1, 2, and 5.  These values are in the ``grey zone''
between the model values for UV pumping and shocks.  However, thermal
collisions may lower these ratios, and the presence of high
$\nu$-level lines ($\nu = 4,5,6,7$) suggests that fluorescence plays
an important role \citep{sni07b}.  \citet{whi05} derived a velocity
dispersion between star clusters of $<10$\,km\,s$^{-1}$ from {\sl
STIS} spectroscopy and concluded that high-velocity cloud-cloud
collisions cannot be the dominating starburst mode in the Antennae.
On the other hand, \citet{wil00} found evidence for cloud-cloud
collisions near the strongest mid-IR peak, but \citet{sch07} concluded
that massive shock heating, either due to supernova remnants or due to
cloud collisions, is not likely the dominant heating source for the
SGMCs in the overlap region.

A starburst system of particular interest for comparison may be
Stephan's Quintet (NGC\,7317/7318b)-- a compact group of four strongly
interacting galaxies with a likely foreground galaxy.  \citet{app06}
reported on the discovery of an almost ``pure H$_2$ line spectrum''
with unusually strong and broad H$_2$ lines, resembling the spectra of
shocked gas seen in Galactic supernova remnants.  The total mass of
warm H$_2$ was determined to be $3.4\cdot 10^7$\Msol, about 40\% more
than what we measured in the Antennae.  \citet{app06} attributed the
powerful H$_2$ emission from Stephan's Quintet to a large-scale shock
wave, resulting from a high-velocity galaxy collision.  The evidence
for this claim is manifold: the unusually broad line width of $\sim
870$\,km\,s$^{-1}$, the relative strength of the H$_2$ S(0) - S(5)
lines as indicated in the excitation diagram (their Fig.~3), and the
absence of PAH-dust features and very low excitation ionized gas
tracers.

So far, the strongest support for large-scale shock excitation in the
Antennae came from {\sl ISO} data \citep{haa05}, which showed
exceptionally strong H$_2$\,S(3) emission from the overlap region.  In
addition, the detected emission appears also spatially displaced from
the known starburst regions (Fig.~\ref{figH2haas}).  \citet{haa05}
interpret these two findings as combined evidence for pre-starburst
shocks that arise from neutral {\sc H\,i} cloud-cloud
collisions.  

As discussed in Section~\ref{secH2strength} our {\sl IRS} observations
disagree with the measurements by \citet{haa05}.  We find that the
H$_{2}$ emission appears to be reasonably well correlated with the
starburst activity in the southern part of the overlap region.  The
only exception may be the nucleus of NGC\,4039 for which -- to our
knowledge -- no published near-IR H$_2$ flux measurements exist, but
the spectra of \citet{gil00} and \citet{men01} suggest that the ratio
is very small and lies within the `shocked' regime.  Based on the
near-IR lines of H$_{2}$\,1-0 S(1) and Br$_{\gamma}$, \citet{fis96}
argued that the surface brightness at the nucleus of NGC\,4039 is too
high to be explained by UV excitation, and favoured C-shocks.

In summary, we find no evidence for large-scale pre-starburst shocks
due to {\sc H\,i} cloud-cloud collisions.  We also note that we have
seen none of the strong evidence for large-scale shocks, that has been
reported for Stephan's Quintet by \citet{app06}, in the Antennae.  In
the overlap region the H$_2$ emission agrees well with the starburst
activity, and leaves both FUV photons and local shocks, e.g. from
supernovae, as origin for H$_2$ excitation.  In the southern nucleus,
from which 45\% of the H$_2$\,S(3) is detected, shocks appear to be
the dominant trigger.  Further observations at higher resolution,
e.g. by {\sl JWST}, are necessary to get a clearer picture of the
complex processes in this interacting system.


\section{Summary} 
\label{secsummary}
We observed the Antennae galaxies (NGC\,4038/39) with the Infrared
Spectrograph on board of the Spitzer Space Telescope to study the
properties of the ISM and the most luminous super-star clusters in
this prototype merger system.  We obtained low-resolution ($R\sim
100$) spectral maps of the entire central area and high-resolution
($R\sim 600$) spectra of six infrared-luminous regions and the two
galactic nuclei.  The high signal-to-noise spectra allow for a detailed
study of fine-structure lines, PAHs, silicates and molecular hydrogen
in the range between $5 - 38\mu$m.

Both nuclei of NGC\,4038 and NGC\,4039 are surrounded by regions of
moderately active star formation.  While the nucleus of NGC\,4039
appears more active in terms of the [\ion{Ne}{3}], [\ion{S}{4}] and
H$_2$\,S(3) emission lines, we can rule out the presence of an AGN in
both nuclei, based on the lack of [\ion{Ne}{5}] emission and the low
[\ion{O}{4}]/[\ion{Ne}{2}] ratio.
Generally, the older super star clusters are associated with softer
radiation, as measured by the [\ion{Ne}{3}]/[\ion{Ne}{2}] ratio.  The
by far hardest and most luminous radiation originates from two compact
clusters (our peaks~1 and2) in the southern part of the overlap
region.  The tracers of softer radiation and PAH emission -- typical
for photo-dissociation regions -- are spatially extended throughout
and beyond the overlap region.

From the slope of the continuum emission we found that peaks~1 and 2
have the highest dust temperatures while peaks~3 and 5 have the
lowest.  However, the differences are subtle and in agreement with the
average spectral slope and diversity of starburst galaxies.
The total infrared luminosity of our six IR peaks plus the two nuclei
is $L_{\rm IR} = 3.82\cdot 10^{10}$\Lsol.  From the IR luminosities we
derived the star formation rates for the individual regions, which
range between 0.2 and 2~\Msol yr$^{-1}$ with a total of 6.6~\Msol
yr$^{-1}$.

We found weak evidence for suppressed PAH emission from regions with
harder ([\ion{Ne}{3}]/[\ion{Ne}{2}]) emission.  However, there is a
clear trend that regions with stronger (i.e., harder {\sl and}
intenser) radiation fields show reduced PAH strength.  Since the
\2~regions and PDR interfaces cannot be spatially resolved at the
distance of the Antennae, aperture effects play an important role in
the interpretation of the results.
We found no evidence for PAH grain size variations from the comparison
of the $17\mu$m and $11.3\mu$m PAH features.  To first order, the PAH
features trace each other very well.  However, we find some evidence
for spatial variations in the $8.6\mu$m \,/\,$11.3\mu$m ratio, which
could be attributed to PAH ionization effects.

The H$_2$ lines are among the strongest emission lines in the
Antennae.  The emission is rather confined to the region around the
nucleus of NGC\,4039 and the southern, more active, part of the
overlap region.  We find a total line flux of $6.1\cdot
10^{-20}$~W\,cm$^{-2}$ in the S(3) line.  This is a factor five less
than previous claims by \citet{haa05} based on {\sl ISOCAM-CVF} data.
With an S(3) line luminosity of $9.2\cdot 10^6$\Lsol\ the luminosity
ratio between the warm molecular gas (as measured by the H$_2$ lines)
to the total FIR emission is $1.6\cdot 10^{-4}$, a value similar to
that found in many starburst and ultra-luminous infrared galaxies.
We did not find the tight correlation between the H$_2$ line and the
PAH strength that was seen by \citet{rou07} on larger scales in the
sample of {\sl SINGS} galaxies.  Apart from the general ``smoothing
effect'' when averaging over larger regions local shocks from
supernovae are likely to contribute to the discrepancy.

We derived a temperature of the warm H$_2$, averaged over the
individual star forming regions, of $302\pm 26$\,K.  PDRs which have
stronger PAH emission relative to the continuum emission have warmer
molecular gas.  Interestingly, the hardness of the radiation field
anti-correlates with the H$_2$ temperature.  We explain this by the
PDR geometry: older clusters had more time to shape their
surroundings; the UV photons can penetrate farther into the cloud and
provide a more uniform temperature distribution, while young clusters
may have hotter gas and dust near their inner shell but appear cooler
at their well shielded outskirts.
We calculated a total mass of warm H$_2$ in the Antennae of $2.5\cdot
10^7$\Msol.  The fraction of warm to total H$_2$ gas mass is 0.35\%
globally and varies between 0.4\% and 0.8\% for the individual
supergiant molecular clouds.

We find no evidence for large-scale pre-starburst shocks due to {\sc
H\,i} cloud-cloud collisions.  Whether the H$_2$ emission is due to
local shocks or UV pumping cannot be unambiguously determined from our
data.  In the overlap region the H$_2$ emission agrees well with the
starburst activity, and leaves both FUV photons and local local
shocks, e.g. from supernovae, as origin for H$_2$ excitation.  In the
southern nucleus, from which 45\% of the H$_2$\,S(3) is detected,
shocks appear to be the dominant excitation mechanism.


\acknowledgments

We would like to thank Dr.\,Zhong Wang, who provided us with the
proprietary {\sl IRAC} images of the Antennae immediately after their
observations, which enabled the optimal planning of our spectroscopic
follow-up.  We also thank Dr.\,Henrik Spoon for stimulating
discussions and Dr.\,Martin Haas for many critical comments, which,
however, did not result in an agreement on the H$_2$
emission. Finally, we would like to thank the anonymous referee for
many comments that helped to clarify the discussion.

This work is based on observations made with the Spitzer Space
Telescope, which is operated by the Jet Propulsion Laboratory,
California Institute of Technology under NASA contract 1407. Support
for this work was provided by NASA through Contract Number 1257184
issued by JPL/Caltech.  VC would also like to acknowledge partial
support from the EU ToK grant 39965.


%
\clearpage

\appendix

\section{Calculation of the H$_2$ Temperatures and Masses}

The H$_{2}$ molecule exists in two states: ortho-H$_{2}$ with parallel
nuclear spins (odd~$J$), and para-H$_{2}$ with anti-parallel spins
(even~$J$).  In local thermodynamic equilibrium (LTE) the
ortho-to-para ratio is 3 for $T > 200$\,K \citep{bur92}. The S(0) and
S(2) lines are from para-H$_{2}$, while the S(1) and S(3) lines are
from ortho-H$_{2}$.  Hence, the line ratios of the same ``species''
are independent of the ortho-to-para ratio while the ratio of two
lines adjacent in wavelength is sensitive to the ortho-to-para ratio
and to the temperature.  Unfortunately we have to use a mixture of
ortho and para lines for observational reasons: a reasonable
comparison can only be done for lines observed with the same {\sl IRS}
module (slit width).  Hence, our analysis of the `lores' maps will
focus on the S(2) and S(3) lines, and the `hires' spectra on the
S(1) and S(2) lines.

In our calculations we assume that the observed line emission is
optically thin, the critical densities $n_{cr}$ of the observed lines
are $\leq 10^3$\,cm$^{-3}$ (this is certainly given for the S(0)
through S(2) lines), and that the populated levels are in LTE with an
ortho-to-para ratio of 3.  The S-notation refers to transitions with
$\Delta J = J_{upper} - J_{lower} = +2$, and with $g_{J} = g_{s}
(2J+1)$ we derive the quantum numbers listed in Table~\ref{tabh2}.
$g_{s}$ is the nuclear statistical weight factor, which is 1 for
para-H$_2$ and 3 for ortho-H$_2$.

\begin{deluxetable}{r c c c r r l }
\centering 
\tabletypesize{\footnotesize} 
\tablecaption{H$_{2}$ molecular constants\label{tab:h2const}}
\tablewidth{0pt} 
\tablehead{
\colhead{$\lambda$ [$\mu$m]} & 
\colhead{} & 
\colhead{$g_s$} & 
\colhead{$J_{up}$} & 
\colhead{$g_J$} & 
\colhead{$E_{up} / k$ [K]} & 
\colhead{$A_{ul}$ \tablenotemark{a} [s$^{-1}$]} 
}
\startdata 
28.219 & S(0) & 1 & 2 &  5 &  509.9 & $2.94\cdot 10^{-11}$ \\
17.035 & S(1) & 3 & 3 & 21 & 1015.1 & $4.76\cdot 10^{-10}$ \\
12.279 & S(2) & 1 & 4 &  9 & 1681.7 & $2.76\cdot 10^{-9}$  \\
 9.665 & S(3) & 3 & 5 & 33 & 2503.8 & $9.84\cdot 10^{-9}$  \\
\enddata
\label{tabh2}
\tablenotetext{a}{The upper-level energies and Einstein $A$ coefficients
               were taken from \citet{ros00} and references
               therein. $g_s$ is the statistical weight factor and
               $g_{J}$ is the level degeneracy of rotational level
               $J$.}
\end{deluxetable}

In order to derive the excitation temperature $T_{ex}$ we apply the
following calculation: the column density $N_i$ for a given transition
$i$ is related to the total column density $N_{tot}$ via the Boltzmann
distribution 
$N_i = g_J N_{tot} \cdot \exp [-E_i / kT] / Q_{H_2}$, 
where $Q_{H_2}$ is the temperature-dependent partition function of
H$_2$, with 
{\footnotesize
\begin{eqnarray}
Q_{H_2} &=& \sum_{J} (2J + 1) g_s e^{-\frac{E_J}{kT}} = \nonumber \\
        &=& 1 \cdot \left( e^{-\frac{0}{T}} 
        + 5 e^{-\frac{509.0}{T}} 
        + 9 e^{-\frac{1681.7}{T}} 
        + 13 e^{-\frac{3474.6}{T}} 
	+ \ldots \right)  
        + 3 \cdot \left(3 e^{-\frac{169.8}{T}}
        + 7 e^{-\frac{1015.1}{T}}
        + 11 e^{-\frac{2503.8}{T}}
        + 15 e^{-\frac{4586.7}{T}}
	+ \ldots \right). \nonumber \\
\end{eqnarray}}
For a temperature range between $250-400$~K the partition
function $Q_{H_2}$ typically varies from 6 to 10.  Hence, the ratio
between column densities for a given transition is
\begin{displaymath}
\frac{N_1}{N_2} = \frac{g_1}{g_2} e^{\frac{E_2 - E_1}{kT}}.
\end{displaymath}
The line intensity $I_i$ in [W\,m$^{-2}$] is related to the column
density via 
$N_i = \frac{4\pi\lambda}{h c}\frac{I_i}{A_i}$,
and the Einstein coefficient $A_i$ describes the spontaneous
emission probability.  Thus the excitation temperature $T_{ex}$ can be
calculated as
\begin{displaymath}
T_{ex} = \frac{E_2 - E_1}{k \ln \left( \frac{g_2}{g_1}
         \frac{\lambda_1}{\lambda_2} \frac{I_1}{I_2} \frac{A_2}{A_1}
         \right)}.
\end{displaymath}

We emphasize that the mid-infrared H$_2$ lines are only sensitive to
the warm ($\ge 100$~K) gas, which is only a small fraction of the
total H$_2$ gas (see Section~\ref{secH2strength}).  In order to derive
the mass of warm H$_2$ we use the following calculation: for an
ortho-to-para ratio of three the total mass of molecular hydrogen is
$M_{H_2} = \frac{4}{3} M_{ortho} = m_{H_2} N_{H_2}$, where $m_{H_2}$
is the molecular mass of hydrogen, and the total number of molecules
is
\begin{displaymath}
N_{H_2} = \frac{4\pi D^2 \lambda I_i}{h c A_{ul} x_J }.
\end{displaymath}
$x_J$ describes the fractional population in the upper level as
$x_J = N_J / N_{tot} = g_J \exp\left[-E_J / ht \right] / Q_{H_2}$.
Combining the above equations we get for the warm H$_2$ mass:
\begin{displaymath}
M_{H_2} = \frac{4\pi m_{H_2} D^2 \lambda I_i
          Q_{H_2}}{h c A_{ul} g_J e^{\frac{-E_J}{kT}}}.
\end{displaymath}


\end{document}